\documentclass[aip,jcp,amsmath,amssymb,floatfix,reprint,citeautoscript,articletitle=false,noeprint]{revtex4-1}

\usepackage[english]{babel}
\usepackage{color}
\usepackage{graphicx}
\usepackage[caption=false]{subfig} 
\usepackage{amsmath,amssymb,bm}
\usepackage[version=3]{mhchem}
\usepackage{verbatim}
\usepackage{multirow}
\usepackage{dcolumn}
\usepackage{float}
\usepackage{nicefrac}
\usepackage{siunitx}
\usepackage{booktabs}
\DeclareSIUnit[number-unit-product = {\,}]{\au}{a.u.}
\DeclareSIUnit[number-unit-product = {\,}]{\kJmol}{\kilo\joule\per\mol}
\usepackage[colorlinks,allcolors=black,citecolor=blue,urlcolor=blue]{hyperref}

\begin{document}
\title{%
Water flow in single-wall nanotubes: Oxygen makes it slip, hydrogen makes it stick
}
\author{Fabian L. Thiemann}
\affiliation{%
Thomas Young Centre, London Centre for Nanotechnology, and Department of Physics and Astronomy, University College London, Gower Street, London, WC1E 6BT, United Kingdom
}%
\affiliation{%
Yusuf Hamied Department of Chemistry, University of Cambridge, Lensfield Road, Cambridge, CB2 1EW, UK
}
\affiliation{%
Department of Chemical Engineering, Sargent Centre for Process Systems Engineering, Imperial College London,
South Kensington Campus, London SW7 2AZ, United Kingdom
}%
\author{Christoph Schran}
\email{cs2121@cam.ac.uk}
\affiliation{%
Yusuf Hamied Department of Chemistry, University of Cambridge, Lensfield Road, Cambridge, CB2 1EW, UK
}
\affiliation{%
Thomas Young Centre, London Centre for Nanotechnology, and Department of Physics and Astronomy, University College London, Gower Street, London, WC1E 6BT, United Kingdom
}%
\author{Patrick Rowe}
\affiliation{%
Yusuf Hamied Department of Chemistry, University of Cambridge, Lensfield Road, Cambridge, CB2 1EW, UK
}
\affiliation{%
Thomas Young Centre, London Centre for Nanotechnology, and Department of Physics and Astronomy, University College London, Gower Street, London, WC1E 6BT, United Kingdom
}%
\author{Erich A. Müller}
\email{e.muller@imperial.ac.uk}
\affiliation{%
Department of Chemical Engineering, Sargent Centre for Process Systems Engineering, Imperial College London,
South Kensington Campus, London SW7 2AZ, United Kingdom
}%
\author{Angelos Michaelides}
\email{am452@cam.ac.uk}
\affiliation{%
Yusuf Hamied Department of Chemistry, University of Cambridge, Lensfield Road, Cambridge, CB2 1EW, UK
}
\affiliation{%
Thomas Young Centre, London Centre for Nanotechnology, and Department of Physics and Astronomy, University College London, Gower Street, London, WC1E 6BT, United Kingdom
}%

\date{\today}

\begin{abstract}
Experimental measurements have reported ultra-fast and radius-dependent water transport in carbon nanotubes which are absent in 
boron nitride nanotubes.
Despite considerable effort, the origin of this contrasting (and fascinating) behaviour is not understood. 
Here, with the aid of machine learning-based molecular dynamics simulations that deliver first-principles accuracy,
we investigate water transport in single-wall carbon and boron nitride nanotubes. 
Our simulations
reveal a large, radius-dependent hydrodynamic slippage on both materials with water experiencing indeed a $\approx 5$ times lower friction on carbon surfaces compared to boron nitride.
Analysis of the diffusion mechanisms across the two materials reveals that the fast water transport on carbon is governed by facile oxygen motion, whereas the higher friction on boron nitride arises from specific hydrogen-nitrogen interactions.
This work not only delivers a clear reference of unprecedented accuracy for water flow in single-wall nanotubes, but also provides detailed mechanistic insight into its radius and material dependence for future technological application.

\end{abstract}
\maketitle

\section{Introduction}
\label{sec:intro}

The ability of water to flow seemingly friction-less across graphitic surfaces \cite{Majumder2005, Holt2006, Secchi2016,Tunuguntla2017,Xie2018,Keerthi2021,Munoz-Santiburcio2021} has put carbon nanotubes (CNTs) at the forefront of nanofluidic \cite{Bocquet2010,Bocquet2020} applications in the fields of desalination\cite{Elimelech2011,Logan2012}, water filtration\cite{Cohen-Tanugi2012,Park2014}, and blue energy harvesting \cite{Siria2017}.
In particular, recent experiments \cite{Secchi2016} in CNTs have
shown that water exhibits an enormous and curvature-dependent hydrodynamic slippage (low friction) with smaller radii resulting in a greater slippage. 
In contrast, in
isostructural but electronically different boron nitride nanotubes (BNNTs)
no slip was detected.
To exploit the full potential of low dimensional materials for nanofluidic devices, a clear understanding of the physical mechanisms behind this radius and material dependence is required \cite{Faucher2019}.

Despite more than a decade of intense research, however, our understanding of 
the transport properties of water inside nanotubes remains far from complete.
This lack of insight partially arises from: (i) differences in the systems studied experimentally (single multi-walled CNTs \cite{Secchi2016a},  carbon nano-conduits \cite{Tunuguntla2017,Xie2018,Keerthi2021}, and membranes of aligned CNTs \cite{Majumder2005, Holt2006}); (ii) the challenge of accurately measuring flow through extremely narrow channels; and (iii) the likely sensitivity of the results to impurities and defects that are inevitably present.
Molecular dynamics (MD) simulations allow, in principle, for these challenges to be bypassed \cite{Muller2013}. 
However, when classical MD simulations have been performed the results obtained are highly sensitive to the interaction models used and computational setups employed; showing a three orders of magnitude spread for the flow enhancement of water inside CNTs \cite{Kannam2013}.
In addition, classical MD simulations fail to explain the experimentally observed radius-dependence in the diameter range between $\approx 30 - 100$~nm \cite{Thomas2008,Falk2010}.
\textit{Ab initio} MD (AIMD), conversely,  could provide the required accuracy by accounting explicitly for the electronic structure of the systems studied \cite{Ruiz-Barragan2019}. 
Indeed AIMD simulations have revealed that water exhibits a $3-5$ times larger friction on hexagonal boron nitride surfaces compared to graphene \cite{Tocci2014a,Tocci2020}.
These studies, however, have been limited to flat sheets as the high computational cost of AIMD impedes the simulation of large diameter nanotubes.
The inherent constraints on the accessible length and time scales, moreover, inevitably introduce finite size errors and question marks over the convergence of the dynamical quantities computed.
Thus, despite the progress made, a systematic study of both the radius and material dependence -- using techniques that accurately tackle the interatomic interactions and dynamical properties -- has yet to be performed.

Here, we rise to this challenge and report the findings of a detailed first-principles machine learning-based molecular dynamics study of water transport in single-wall CNTs and BNNTs. 
The key aims of this work are: (i) to obtain reliable reference-quality first principles values for water flow, and in so-doing shed light on the myriad of simulation results in the literature  \cite{Kannam2013}; and (ii) to gain molecular-level understanding of the mechanisms of water transport in low-dimensional materials.
By reliably representing the potential energy surface (PES) of a chosen first principles reference method, machine learning potentials (MLPs) have become a powerful approach for simulating complex systems, achieving quantum mechanical accuracy at a fraction of the usual cost and, thus, facilitating simulations at longer time and length scales \cite{Behler2016, Deringer2019, Deringer2021,Behler2021}.
Employing our recently introduced methodology for the rapid development of machine learning potentials (MLPs) \cite{Schran2021} allows us to do precisely this, thereby achieving converged statistics while maintaining first-principle accuracy.
The simulation lengths and systems sizes of this work go beyond previous AIMD studies by at least an order of magnitude with more than 40 ns of high-quality simulation data obtained on  nanotubes varying in diameter between $\approx 1.6$ and $\approx 5.5$~nm. 
This allows for the first time to provide a clear reference of unprecedented accuracy for  water flow in single-wall nanotubes.

In agreement with experiments \cite{Keerthi2021} and previous AIMD studies\cite{Tocci2014a,Tocci2020}, we find that water indeed experiences a significantly larger friction in BNNTs compared to CNTs.
The strong curvature-dependence, conversely, is by no means unique to the water-carbon couple but also occurs in BNNTs.
Beyond providing a firm theoretical foundation for flow through pristine single-wall nanotubes, our simulations allow insight into the elementary processes involved.
Specifically, we find that the differences between the two materials originates from alternating docking and hopping events induced by the hydrogen-nitrogen interaction only present in BNNTs (hydrogen-imposed).
The radius dependence observed, conversely, is mainly of geometric nature where a higher curvature results in a smoother free energy landscape, i.e. lower energy barriers, and, thus, smaller friction (oxygen-imposed).
Having a clear understanding of these mechanisms is expected to be of great importance for materials design of nanofluidic devices suggesting routes for directional flow via tailor-made nanotubes or 2D nanostructures.
In this way, our work pushes forward our understanding of water transport under confinement and helps to close a long-standing knowledge gap \cite{Faucher2019} in the field of nanofluidics.
\section{Determination of the material and radius-dependent friction based on first-principles quality machine learning potentials}
\label{sec:results}

\begin{figure*}[t]
    \centering
    \includegraphics[width=0.9\textwidth]{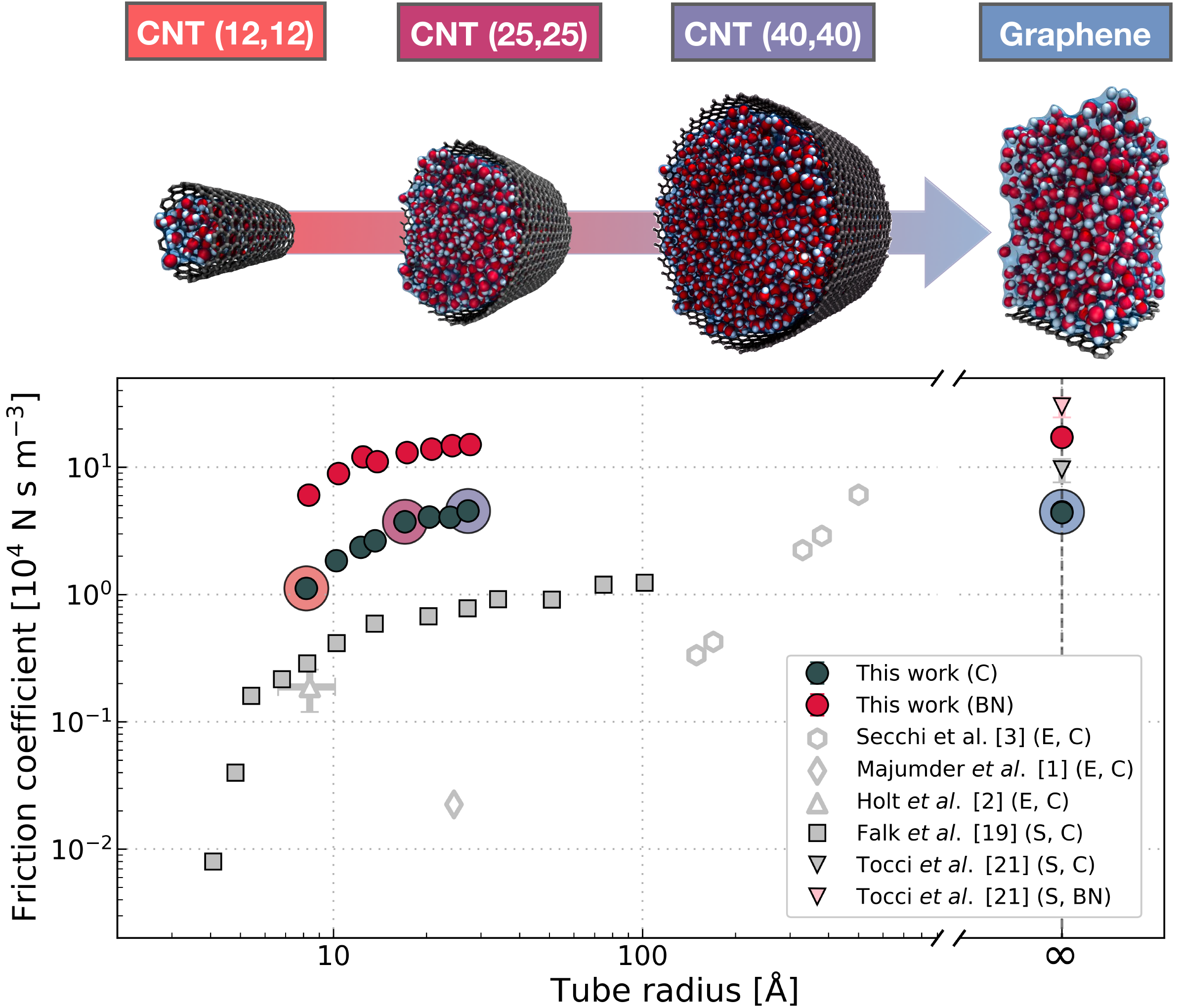}
    
    \caption{\textbf{Friction of water inside CNTs and BNNTs of different diameters.}
    The top panel shows snapshots of the simulations of the selected CNTs and graphene with increasing diameter from left to right.
    In the bottom panel we report the friction coefficient as a function of tube radius showing our results as well as a small selection of previous experimental and computational work.
    Depending on the type of study, the related data is labeled with E (experiment) and S (simulation), respectively.
    Similarly, the confining material investigated is indicated by C (CNTs and graphene) and BN (BNNTs and hBN).
    The circles around the data points in the lower panel correspond to the systems shown in the top panel with the corresponding colour.
    From our simulations, the statistical error was obtained from splitting the trajectory into two blocks.
    However, the magnitude of the error is small compared to the marker size on the log-log scale.
    \label{fig:friction}}
\end{figure*}

Using the approach introduced in \cite{Schran2021} we developed and validated MLPs to probe the systems targeted in this study. Details of the approach used and validations are provided in the Methods section and the SI. 
With these MLPs, we proceed to benchmark the hydrodynamic slippage of water inside single-wall nanotubes.
The friction coefficient $\lambda$ can be directly computed from equilibrium MD simulations using a well known Green-Kubo relationship \cite{Bocquet1994}.
In figure \ref{fig:friction} we show the dependence of $\lambda$ on the tube diameter computed in this work for 16 different nanotubes as well as graphene and h-BN surfaces. 
Also shown is a -- by no means comprehensive -- selection of results obtained in previous work to illustrate the wide spread of results, 
which we will address in detail below.

Based on our simulations, we find that irrespective of the curvature, water exhibits a $\approx 4-5$ times larger friction coefficient on BN surfaces compared to equivalent carbon systems reaching a maximum value of $\approx 4.5\cdot 10^4$~N~s~m$^{-3}$ and $\approx 17\cdot10^4$~N~s~m$^{-3}$ for monolayer graphene and hBN, respectively.
These friction coefficients on the curvature-free interfaces agree well with previous computational studies \cite{Falk2010,Tocci2014a,Tocci2020,Rajan2019,Ghorbanfekr2020,Poggioli2021}.
In fact, our benchmark simulations provide a reliable estimate of the absolute values which are highly scattered ranging from $\approx 1 $ to $\approx 10 \cdot 10^4$~N~s~m$^{-3}$ (experiments \cite{Maali2008} report a friction coefficient of $\approx 12 \cdot 10^4$~N~s~m$^{-3}$ on graphite) and $\approx 4$ to $\approx 30 \cdot 10^4$~N~s~m$^{-3}$ for the distinct systems.
This wide spread of results can be associated with differences in the chosen force field\cite{Oga2019,Rajan2019,Poggioli2021}, DFT functional\cite{Tocci2014a,Tocci2020}, or simulation setup related to a frozen substrate \cite{Falk2010}, finite size errors, thermostating \cite{Sam2017} as well as confinement of water between two layers \cite{Ghorbanfekr2020, Tocci2020}.

\begin{figure*}[t]
    \centering
    \includegraphics[width=\textwidth]{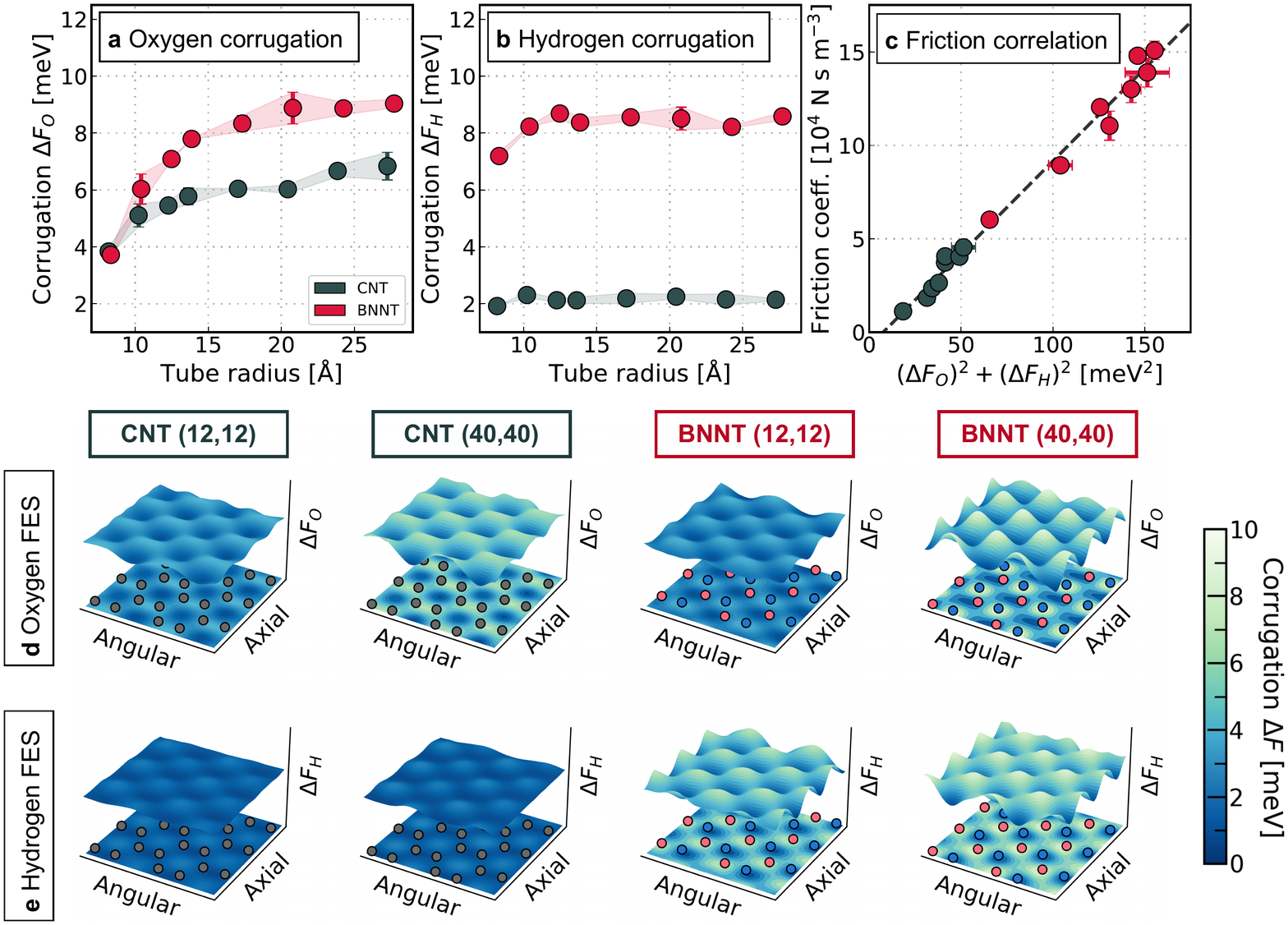}
    \caption{\textbf{Linking the friction to the free energy surface of water confined to CNTs and BNNTs.}
    \textbf{Part a}: Corrugation $\Delta F_O$ of the oxygen-based free energy surface (FES) for CNTs and BNNTs plotted as function of the tube radius. 
    The error bars correspond to the statistical error which was obtained by splitting each trajectory into two blocks.
    \textbf{Part b}: Corrugation $\Delta F_H$ of the hydrogen-based FES for CNTs and BNNTs plotted as function of the tube radius.
    \textbf{Part c}: Correlation between the friction coefficient and the sum of the squared corrugations.
    The dashed line represents a linear fit to the data obtained via orthogonal distance regression.
    \textbf{Part d}: Visualisation of the oxygen-based FES for the smallest and largest CNTs and BNNTs.
    The solid atoms are represeted by the markers in the projection where carbon, boron, and nitrogen are coloured in grey, pink, and blue.
    \textbf{Part e}: Visualisation of the hydrogen-based FES for the smallest and largest CNTs and BNNTs.
    \label{fig:fes}}
\end{figure*}
In nanotubes, for both materials a stark radius dependence is observed where smaller diameters lead to a significantly reduced friction of $\approx 1\cdot 10^4$~N~s~m$^{-3}$ and $\approx 6\cdot10^4$~N~s~m$^{-3}$ for the smallest CNT and BNNT (radius $\approx 0.8$~nm), respectively.
As an illustration of the dimension of this effect, we highlight that the friction inside the smallest BNNT approaches the value of graphene which is generally considered to exhibit a large hydrodynamic slippage.
For larger tube diameters, the friction coefficient converges to the value of the flat surface for both materials at radii $\gtrapprox 2.5$~nm. 
This first-principle estimate is one order of magnitude smaller than observed in experiments \cite{Secchi2016} and rather similar to findings of previous force-field based simulations \cite{Falk2010}. 
In fact, the friction in CNTs predicted by our simulations generally exceeds the %
values obtained in nanofluidic measurements in isolated \cite{Secchi2016} and membranes of multi-wall \cite{Majumder2005,Holt2006} CNTs.
For BNNTs, moreover, we observe slippage of considerable extent opposed to the experiments \cite{Secchi2016}.
We will discuss these deviations between experiments and simulations in detail in section \ref{sec:discussion}.
For now, however, we focus on understanding the physical mechanisms behind the radius and material dependence observed in our reference simulations.

\section{Unveiling the distinct roles of oxygens and hydrogens in water transport}
Solid-liquid friction is strongly determined by the (free) energy barriers that molecules have to overcome to move across the surface. 
Thus, we begin by examining the free energy surface (FES) of the water molecules in the contact layer to further understand the radius and material dependent slippage.
In particular, we investigate the overall corrugation of the FES with its square being proportional to the friction coefficient \cite{Barrat1999, Bocquet2010}, such that $\lambda \propto (\Delta F)^2$.
In previous work \cite{Ho2011,Falk2010, Falk2012,Tocci2014a,Tocci2020} the analysis of the potential and free energy profiles has been limited to the oxygen atoms of the water molecules.
With a recent study \cite{Poggioli2021} suggesting that the material dependence could be attributed to hydrogen-nitrogen interactions, here we examine the free energy barriers for both the oxygens and the hydrogens separately. 

In figure \ref{fig:fes} we show how the FES of hydrogens and oxygens varies between materials and with curvature.  
To this end, we illustrate selected FESs for the smallest and largest CNTs and BNNTs investigated and plot the corrugation as a function of the tube diameter.
Focusing on the oxygen corrugation (part a) and profiles (part d) at first, it is clear that the FES becomes more corrugated with increasing radius. 
The energetically favourable positions of the oxygens in the contact layer, conversely, do not vary with curvature and coincide with those observed on flat surfaces \cite{Tocci2014a}. 
On carbon surfaces, oxygen atoms preferentially sit on the hollow site in the middle of a hexagon of carbon atoms.
At the BN interface, in addition to the hollow site, oxygen atoms show an additional free energy minimum around the boron atom. 
The minima observed agree with previous DFT \cite{Al-Hamdani2015} and DMC calculations for the flat graphene and h-BN sheets \cite{Al-hamdani2017,Brandenburg2019a}.

Although the smoothening of the oxygen FES can qualitatively explain the radius-dependence inside single-wall nanotubes, showing an almost identical corrugation for the smallest CNT and BNNT it cannot justify a 5-times larger friction.
\begin{figure*}[t]
    \centering
    \includegraphics[width=\textwidth]{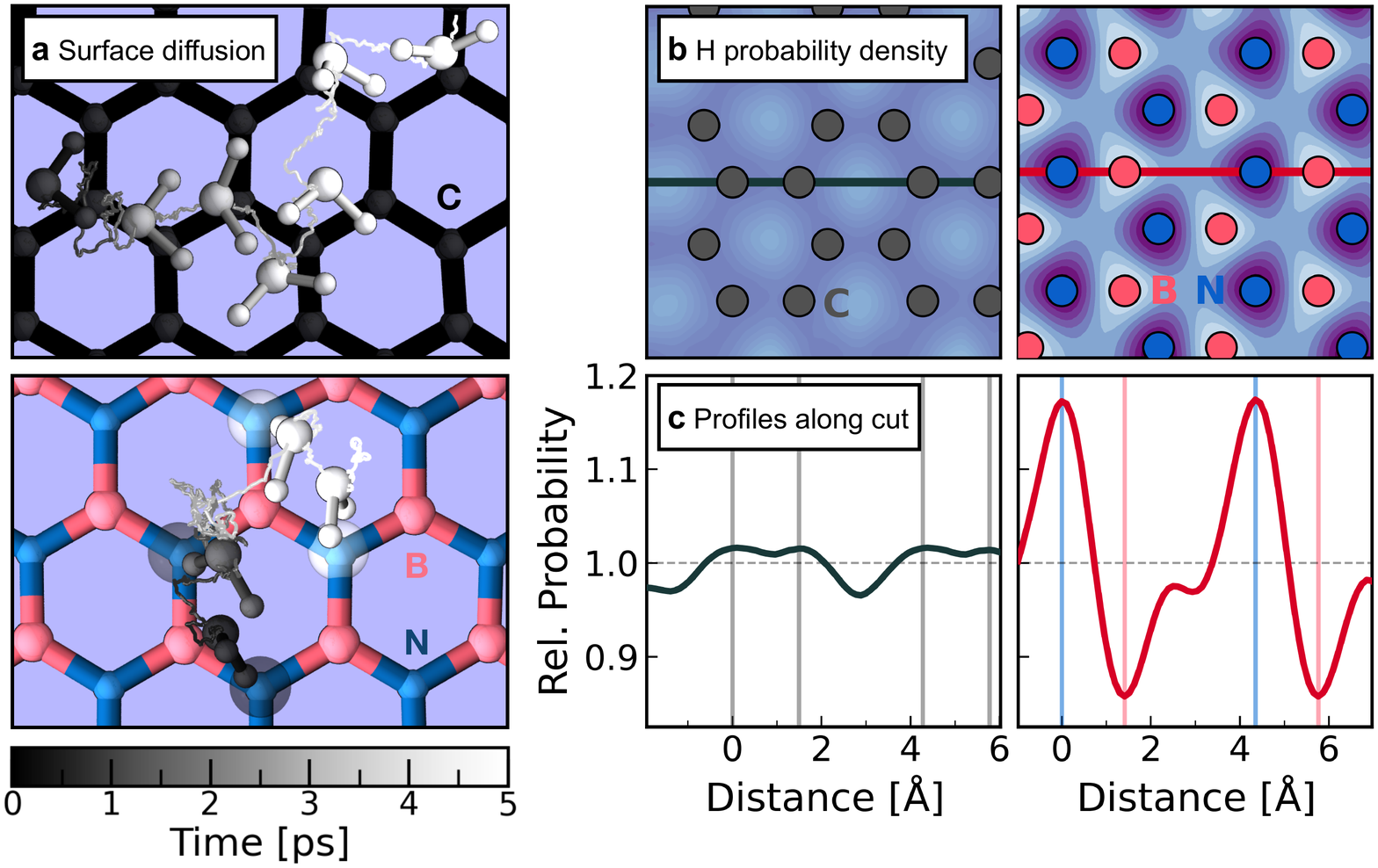}
    \caption{\textbf{Transport mechanisms of water across carbon and BN surfaces.}
    \textbf{Part a:} Snapshots of the trajectory of an individual water molecule in the contact layer diffusing across graphene (top) and hBN (bottom).
    Both the path and individual snapshots of the water molecule are colour-coded according to the time spanning overall 5 ps.
    In the bottom panel, the respective nitrogens involved in the docking events are coloured according to the color of the water molecule at the given time.
    \textbf{Part b:} Two-dimensional probability density of the hydrogen atoms in the contact layer on graphene (left) and hBN (right).
    For both materials, we use the identical scale of the color-coding where low and high probability correspond to light blue and dark purple, respectively.
    The coloured markers represent the average position of the solid atoms and the lines illustrate where the probability density is cut along for further analysis.
    \textbf{Part c:} Profiles of the two-dimensional probability density along the cut directions for graphene (left) and hBN (right). 
    The probability is expressed relative to the average probability of the respective system.
    The vertical lines represent the average position of the solid atoms shown in the panel above.
    \label{fig:mechanism}}
\end{figure*}
In stark contrast to the oxygen atoms, the hydrogen-based FES features only a very weak radius dependence for both materials as shown in part b and e of figure \ref{fig:fes}.
Moreover, there is a pronounced difference between carbon and BN interfaces with the latter showing a roughly 4 times higher corrugation for the hydrogen FES.
Interestingly, for the smallest BNNT the corrugation of the hydrogen-FES is almost twice as large as that of the oxygen FES.
Unsurprisingly, the hydrogen atoms preferably adopt the positions not occupied by the oxygens.
In CNTs this refers to the positions around the carbon atoms while the free energy minimum is around the nitrogen atoms in BNNTs.
The interaction between hydrogens and nitrogens in BNNTs is too weak to be classified as hydrogen bonding.
However, it still yields an enhanced barrier for the water molecule to overcome, which explains the difference in friction observed for the smallest nanotubes.
Here, for the first time, we are able to quantify the magnitude of this effect by comparing it to the non-polar carbon surfaces and show that it is almost independent of the curvature of the confinement.
Further, we show in part e of figure \ref{fig:fes} that there is a linear relation between the friction coefficient and the sum of the squared corrugation;  highlighting the importance of accounting for the contributions from both oxygen and hydrogen.

The results of our separate analysis of the FES point towards a distinct motion pattern on both surfaces, explaining the significantly larger friction in BNNTs compared to CNTs. 
To understand this novel mechanism we follow the trajectory of individual water molecules in the contact layer next to the solid surface.
Part a of figure \ref{fig:mechanism} illustrates this on the flat graphene and hBN sheet for a selected time period of 5~ps.
As illustrated by the FES, the transport of water on graphene is mainly determined by the position of the oxygen with the orientation of the molecule being relatively unimportant. 
This rather unconstrained motion enables fast transport. 
On a hBN surface, conversely, the hydrogens play an important role and govern the diffusion path of the water molecule, as highlighted by the corrugation of the FES.
Tracing individual water molecules, we observe a docking mechanism with water adopting a specific configuration (a so-called one leg structure \cite{Al-Hamdani2015}) which fluctuates closely around the nearest nitrogen atom.
The transport across the surface is then characterised by hopping events, where the water molecules perform jumps between nitrogen sites where they then have a longer residence time.

To provide further evidence of the identified hopping-docking diffusion scheme on BN surfaces, we show the two-dimensional probability density of the hydrogen atoms in the contact layer for graphene and hBN in part b of figure \ref{fig:mechanism}.
In contrast to an almost homogeneous distribution on graphene (left panel), the hydrogens preferably arrange above the nitrogen atoms on hBN.
We now compare the profiles of the relative probability along a cut through the density as shown in part c of figure \ref{fig:mechanism}.
With this profile being linked to the transport mechanism, we indeed find a strongly corrugated pattern for hBN corresponding to an impeded reorientation.
These findings also agree well with previous work \cite{Kayal2017} where it was shown that water hydrogens approach nitrogen atoms in hBN considerably closer than the boron atoms or carbon atoms on graphene.
This underlines the observed trends in the trajectories and indicates that this hydrogen-nitrogen interaction is indeed the culprit behind the material dependence.
While the oxygen-driven diffusion of water molecules in CNTs enables a large hydrodynamic slippage, water transport in BNNTs is hydrogen-dominated resulting in a higher friction.

\section{Bridging the gap between simulations and experiments}\label{sec:discussion}

Having provided a clear and consistent picture for water transport in single-wall CNTs and BNNTs, we now return to discuss the evident deviations between our simulations and the nanofluidic measurements \cite{Majumder2005, Holt2006,Secchi2016} (figure \ref{fig:friction}). 
In principle, these differences could stem both from effects not taken into consideration in our simulations and in the experiments.
Starting with the discrepancies found for CNTs, one particularly interesting issue is the potential importance of 
a non-Born-Oppenheimer-based quantum friction that may play an important role in  multi-walled CNTs \cite{Kavokine2021}. 
Specifically, it was suggested that this additional friction is induced by coupling of charge fluctuations in the water to the electronic excitations in the solid.
With the electrons being able to tunnel between stacked layers, this additional term dominates water transport on graphite and multi-walled CNTs of large diameter where individual layers interact strongly \cite{Endo1997}.
At smaller diameters, conversely, the weakening of the interlayer coupling results in a decreasing contribution of this quantum friction which then becomes negligible in single-wall CNTs and graphene.
If quantum friction plays a significant role in multi-wall nanotubes, 
then differences between our simulations on single-wall nanotubes and experimental measurements on multi-wall nanotubes are to be expected. 
A second factor worth taking into consideration is the rigidity of the nanotubes and how this changes with radius and/or upon going from single-wall to multi-wall nanotubes. 
Our simulations reveal a significant difference in the friction between frozen and flexible CNTs (see supplementary information and reference \cite{Sam2017}). 
Should the tube's rigidity increase due to the enhanced interlayer coupling at larger diameter then this could also significantly alter the friction; thus providing a classical explanation for the radius dependence in multi-wall nanotubes.
Going forward, it would therefore be interesting to explore multi-walled nanotubes with the ML framework exploited here as well as attempting to account for the  non-Born-Oppenheimer electronic friction. 
In addition, experimental measurements for single-wall nanotubes and graphene would be particularly welcome. 
Moving now to BNNTs, which are considerably less slippy than CNTs.  
Experiments \cite{Secchi2016} report a slip length of $< 5$~nm for all BNNTs, providing a lower limit to the friction coefficient of $\approx 20$~N~s~m$^{-3}$.
This agrees well with our findings for the large nanotubes and remaining discrepancies could stem from the high surface charge inside BNNTs observed in  experiments \cite{Siria2013,Secchi2016a}.
Recent computational studies based on DFT\cite{Grosjean2016} and AIMD simulations\cite{Grosjean2019} attribute this to the ability of hydroxide ions to bind to boron atoms.
In highly alkaline water (high pH), the large number of chemisorbed ions on the surface could then impede the fluid transport by strongly interacting with the water molecules.
Although we did not observe any dissociation of water molecules in our extensive simulations, the surface charge could be enhanced by defects in the confining material promoting dissociation and, thus, increasing the friction \cite{Joly2016}.
While further investigations on the impact of pH and defects on the friction are required to determine the origin of the lack of flow in BNNTs, our simulations represent an important reference for the pristine surfaces indicating no sign of dissociation.
These findings put stress on the experiments \cite{Secchi2016} and underline the importance of extending the set of nanofluidic measurements in nanotubes.

\section{Conclusion}

In conclusion, we have reported an extensive set of results from first principles based ML potentials on the material and radius dependent friction of water in single-wall nanotubes. 
To obtain a reliable description of water transport on low-dimensional materials, 
we developed a new set of MLPs enabling us to simulate large diameter nanotubes at first-principles accuracy.
We find that the hydrodynamic slippage strongly depends on curvature for both materials and that there is a $\approx 5$ times lower friction coefficient on carbon compared to BN. 
While differences from experiments remain, it is important to note that our benchmark data is based on pristine single-wall nanotubes while the nanofluidic measurements were conducted in multi-walled and potentially defective nanotubes.
By giving reliable values for water transport in defect-free single-wall nanotubes, our work represents a solid foundation to thoroughly understand  hydrodynamic slippage while highlighting the lack of and need for additional experiments.

Beyond providing well-defined reference data, by achieving quantum mechanical accuracy our simulations provide detailed insight into the origin behind the radius and material dependence of the water transport.
To this end, we computed the free energy profile - separately for oxygen and and hydrogen atoms - and find that the radius dependence of the friction is accompanied by a smoothing of the oxygen-based free energy surface with decreasing tube diameter reducing the energy barriers impeding fast transport.
The sticky behaviour of water on BN surfaces, conversely, can be traced back to their distinct chemistry and polarity impacting mostly the hydrogen atoms:
While hydrogens experience low energy barriers when water diffuses across a carbon surface, the hydrogen-based FES on BN surfaces is more corrugated and heterogeneous.
Governed by the hydrogen-nitrogen interaction, the water molecules adapt an alternating hopping-docking motion inside BNNTs translating into a larger friction compared to CNTs.
By linking the transport behaviour of water to this novel mechanism at the nanoscale, our work highlights the importance of the electronic structure of the substrate and provides an explanation of the radius and material dependence in pristine single-wall nanotubes. 
This clear knowledge of the mechanism behind the materials and radius dependence of water flow in nanotubes is expected to enable the design of tailor-made nanofluidic devices for directional flow or blue energy harvesting.

\section{Methodology}

\subsection*{Machine learning potentials}
In this work, we build on the pioneering work of Behler and Parrinello \cite{Behler2007,Behler2021} and follow our recently introduced machine learning framework \cite{Schran2021} to develop and carefully validate committee neural network potentials (C-NNPs) \cite{Schran2020} for the water-carbon and water-BN systems, respectively.
We train our potentials to energies and atomic forces obtained from DFT calculations within the generalised gradient approximation using the dispersion corrected functional revPBE-D3 \cite{Zhang1998,Grimme2010,Grimme2011}.
It is important to note that this level of theory has been shown to accurately reproduce both the experimentally measured structure and dynamics of liquid water \cite{Morawietz2016,Gillan2016,Marsalek2017} as well as the interaction energies of water on graphene and inside CNTs obtained using more advanced methods such as diffusion Monte Carlo (DMC) and coupled cluster theory \cite{Brandenburg2019b}.
To ensure the applicability of our MLPs for all radii investigated, the configurations included in the training set range from bulk water and interfaces with zero curvature to highly confined water in nanotubes.
All models have been trained using the open-source package N2P2 \cite{Singraber2019}

\subsection*{Molecular dynamics simulations}

All MD simulations were performed using the CP2K \cite{Kuhne2020} simulation package at a temperature of 300~K in the NVT ensemble.
The temperature was kept constant using stochastic velocity rescaling thermostats\cite{Bussi2007}, with separate thermostats for the solid and the liquid. 
To account for the coupling between the phonon modes of the confining material and the water vibrations\cite{Ma2015, Marbach2018}, all atoms were treated as flexible.
Dependent on the material and curvature, the system size varied between $\approx 960$ and $\approx 8300$ atoms.
The number of water molecules inside the nanotubes was chosen so that the density was 1.0~g/cm$^3$ corresponding to that of bulk water.
For the graphene and hBN sheets, the water film height was roughly 35 $\mathrm{\AA}$.
The simulation length varied with the number of atoms, however, for all systems investigated a minimum sampling time of 1~ns was achieved.
In total, more than 40 ns of first-principles ML data has been obtained for 18 systems (16 nanotubes).
In addition, by performing an extensive set of rigorous tests and convergence checks we  ensure that our results and the main conclusions are robust with respect to system size effects, 
and the length of the dynamical trajectories.
Furthermore, by investigating the impact of the chosen DFT functional, water density and nuclear quantum effects (NQEs) we find that while absolute numbers might change, the relative trends observed are sustained.
See the supporting information (SI) for details of these tests.

\begin{acknowledgments}
We thank Gabriele Tocci and Laurent Joly for their valuable feedback and fruitful discussions.
We are grateful to the UK Materials and Molecular Modelling Hub for computational resources, which is partially funded by EPSRC (EP/P020194/1 and EP/T022213/1).
Through our membership of the UK’s HEC Materials Chemistry Consortium, which is funded by EPSRC (EP/L000202 and EP/R029431), this work used the ARCHER and ARCHER2 UK National Supercomputing Service ({http://www.archer2.ac.uk}).
We are also grateful for the computational resources granted by the UCL Grace High Performance Computing Facility (Grace@UCL), and associated support services. 
CS acknowledges financial support from the Alexander von Humboldt-Stiftung.
\end{acknowledgments}

\section*{Supplementary material}

\subsection*{Supplementary Information}

In section S1 we provide a comprehensive overview of the methodology and computational details:
the system setup and settings used in the MD simulations (S1.A), the development and validation of the MLPs (S1.B), as well as the computation of properties such as the friction coefficient and the FES (S1.C).
Section S2 discusses how the friction coefficient is affected by certain aspects of the simulation and model.
This involves the impact of the water density (S2.A), the system size (S2.B), the simulation time (S2.C), the chosen DFT functional (S2.D), nuclear quantum effects (S2.E), and the flexibility of the confining material (S2.F).

\subsection*{Supplementary Data}
To illustrate the different motion patterns of water molecules on graphene and hBN we provide videos (V1 and V2) tracing the individual water molecule shown in figure 3.

%
%

%
\end{document}


\maketitle
\onehalfspacing
In this supplementary information, we provide additional details on certain aspects of the study reported in the manuscript. This includes a detailed summary of the computational details and methods as well as an analysis of the sensitivity of the results.
\tableofcontents

\section{Computational details}
\subsection{Molecular dynamics simulation}
\subsubsection{System setup}
All systems considered in this work were simulated in orthorombic simulation cells employing periodic boundary conditions in the axial direction ($z$) for nanotubes and along the in-plane directions ($x$, $y$) for the sheets of graphene and hBN.
We studied diameters ranging from $\approx 16$~ $\mathrm{\AA}$ to $\approx 55$~ $\mathrm{\AA}$ corresponding to system sizes between $\approx 900$ and $\approx 8300$ atoms. 
Irrespective of the diameter, we used 12 axial repetitions for each nanotube yielding a consistent length of $\approx 30$~$\mathrm{\AA}$ in axial direction.
The sheets, conversely, were represented by a 6x10 supercell with a length of the in-plane dimensions of  of $\approx 25$~$\mathrm{\AA}$.
All nanotubes were of armchair chirality ($m$,$m$) where $m$ takes values between 12 and 40.
Previous force-field-based simulations \cite{Sam2019a} have shown that water experiences the lowest friction inside carbon nanotubes of this chirality compared to zigzag ($m$,0) and chiral ($m$, $n \neq m$) systems.
This upper boundary for the hydrodynamic slippage makes armchair nanotubes the ideal reference systems to investigate the material and radius dependence of the water friction. 
The water density inside the nanotubes was set to the bulk limit of  $1.0$~g/cm$^3$.
On the graphene and hBN sheet, conversely, the height of the liquid water film was $\approx 35$~$\mathrm{\AA}$ being almost twice as large compared to a previous \textit{ab initio} molecular dynamics (AIMD) study \cite{Tocci2014a} corresponding to a total of $\approx 2500$~atoms.
For the sake of clarity, we provide an overview of the systems investigated in the manuscript in table \ref{tab:si_overview} containing information about the number of atoms, number of water molecules, and simulation time.

\def\arraystretch{1.5}
\begin{longtable}{p{0.2 \textwidth} p{0.2 \textwidth} p{0.2 \textwidth} p{0.3 \textwidth}}
\toprule
\multirow{2}{*}{\textbf{Topology}}  &
 \multicolumn{2}{c}{\textbf{Simulation details}} &
 \multicolumn{1}{c}{\multirow{2}{*}{{\textbf{Illustration}}}} \\
 \cmidrule{2-3}
 & Carbon surface & BN surface &\\
 \hline
 \multirow{4}{*}{Nanotube (12,12)} &R$_\mathrm{tube} = 0.82$~nm
             & R$_\mathrm{tube} = 0.83$~nm& \multirow{4}{*}{\includegraphics[width=0.15\textwidth]{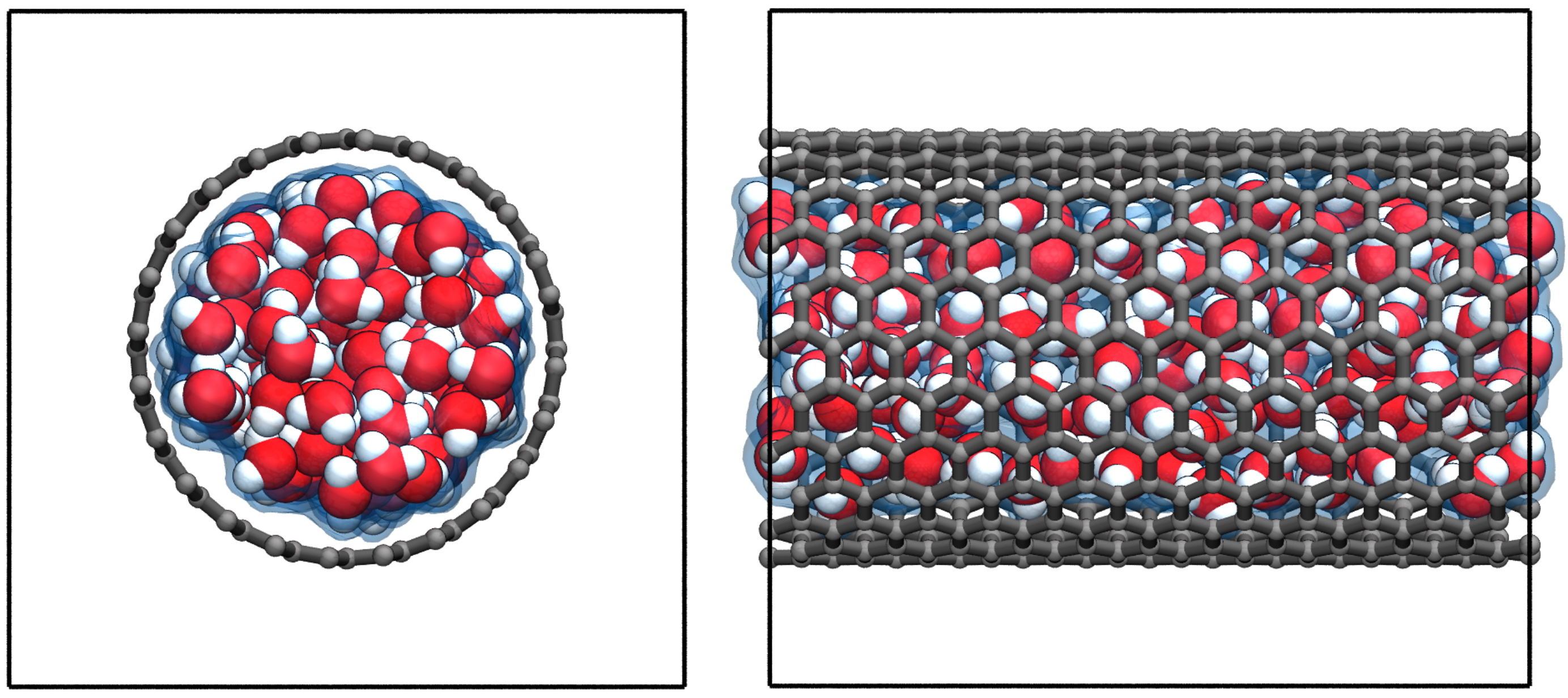}}\\
             & N$_{\mathrm{atoms}} = 966$  & N$_{\mathrm{atoms}} = 984$ & \\
             & N$_{\mathrm{water}} = 130$  & N$_{\mathrm{water}} = 136$ & \\
             &t$_{\mathrm{sim}} = 5$~ns & t$_{\mathrm{sim}} = 5$~ns & \\
             \cmidrule{1-4}
             \multirow{4}{*}{Nanotube (15,15)} &R$_\mathrm{tube} = 1.02$~nm
             & R$_\mathrm{tube} = 1.04$~nm& \multirow{4}{*}{\includegraphics[width=0.17\textwidth]{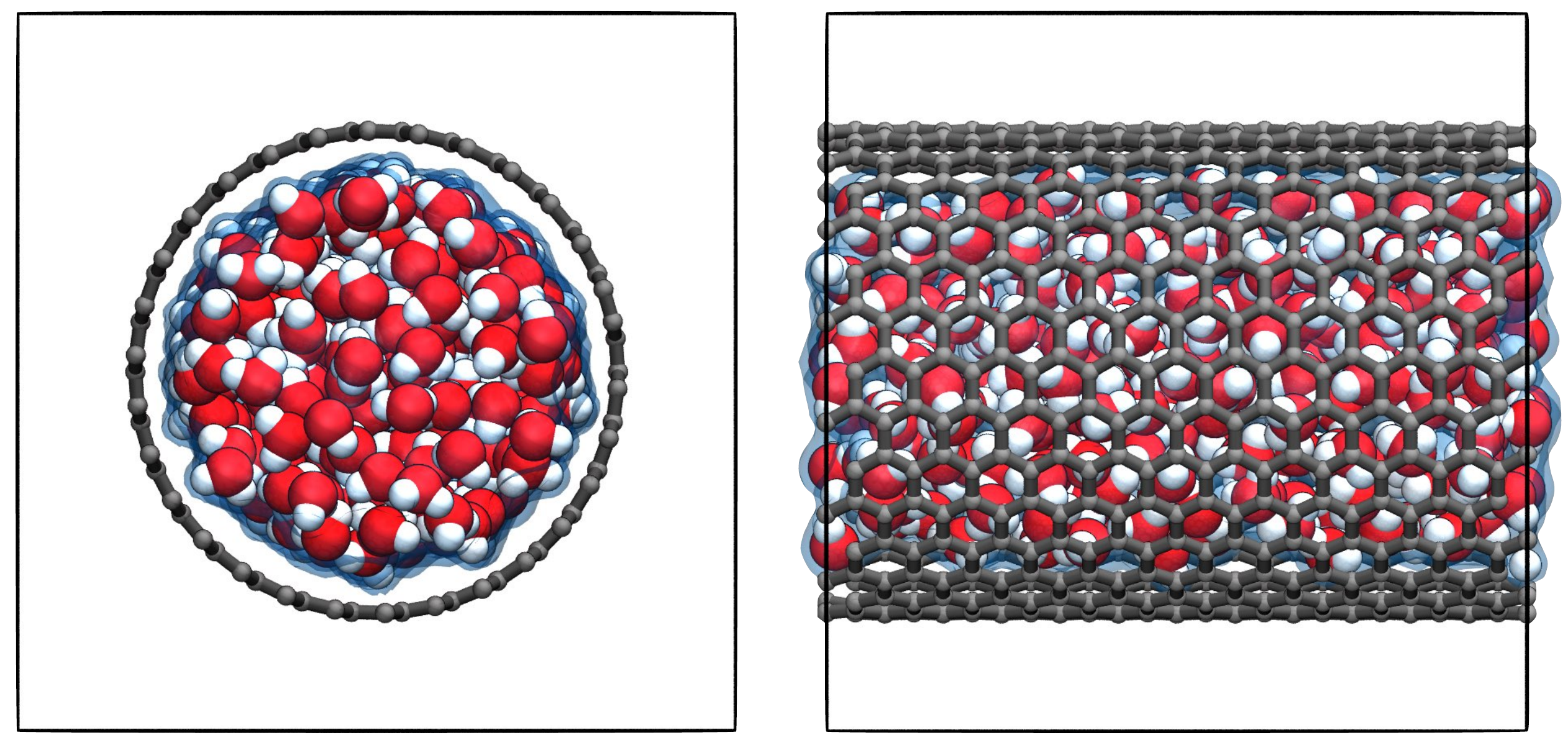}}\\
             & N$_{\mathrm{atoms}} = 1395$  & N$_{\mathrm{atoms}} = 1434$ & \\
             & N$_{\mathrm{water}} = 225$  & N$_{\mathrm{water}} = 238$ & \\
             &t$_{\mathrm{sim}} = 2$~ns & t$_{\mathrm{sim}} = 2$~ns & \\
                \cmidrule{1-4}
             \multirow{4}{*}{Nanotube (18,18)} &R$_\mathrm{tube} = 1.23$~nm
             & R$_\mathrm{tube} = 1.25$~nm& \multirow{4}{*}{\includegraphics[width=0.2\textwidth]{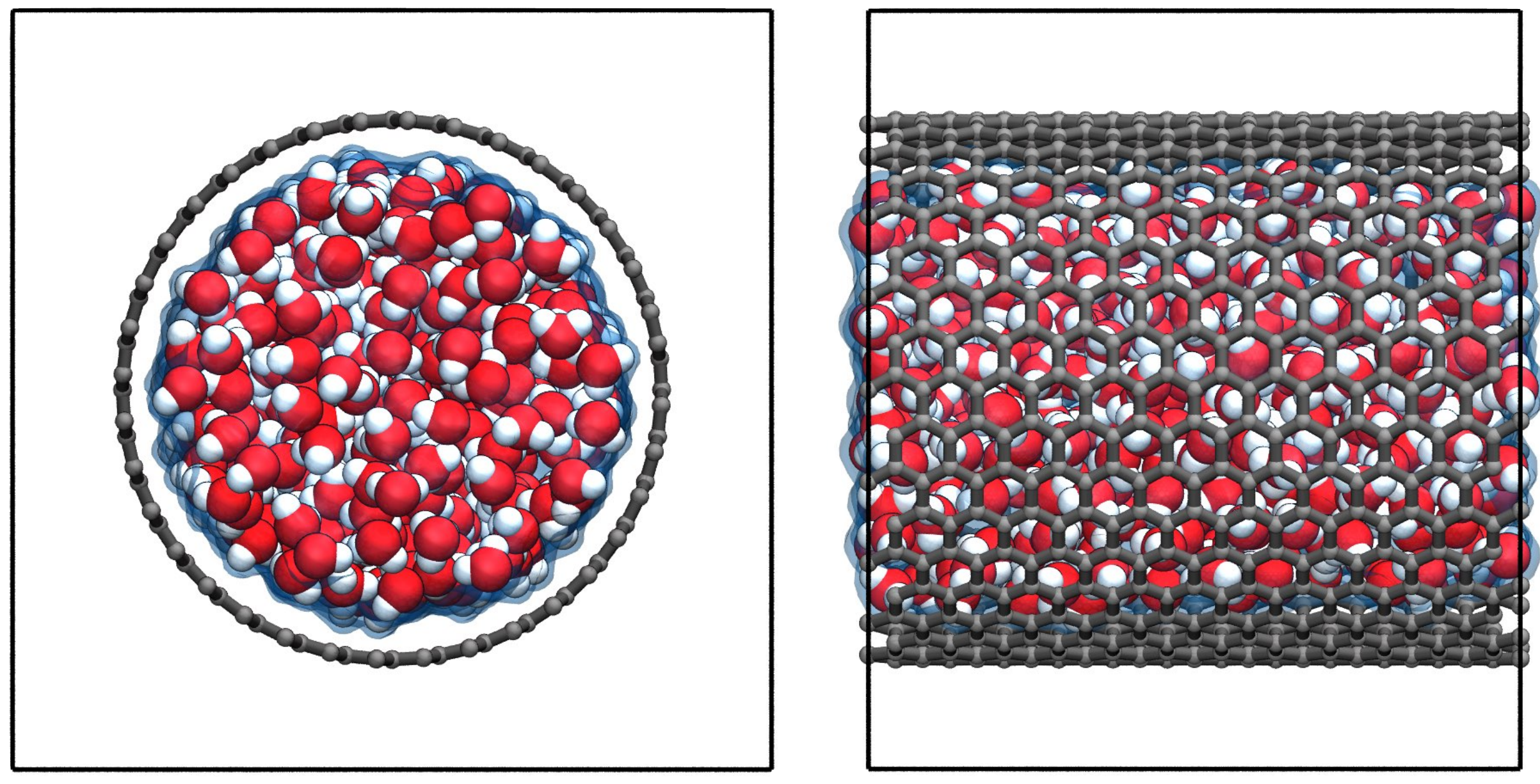}}\\
             & N$_{\mathrm{atoms}} = 1905$  & N$_{\mathrm{atoms}} = 1962$ & \\
             & N$_{\mathrm{water}} = 347$  & N$_{\mathrm{water}} = 366$ & \\
             &t$_{\mathrm{sim}} = 2$~ns & t$_{\mathrm{sim}} = 2$~ns & \\
             \cmidrule{1-4}
             \multirow{4}{*}{Nanotube (20,20)} &R$_\mathrm{tube} = 1.36$~nm
             & R$_\mathrm{tube} = 1.38$~nm& \multirow{4}{*}{\includegraphics[width=0.22\textwidth]{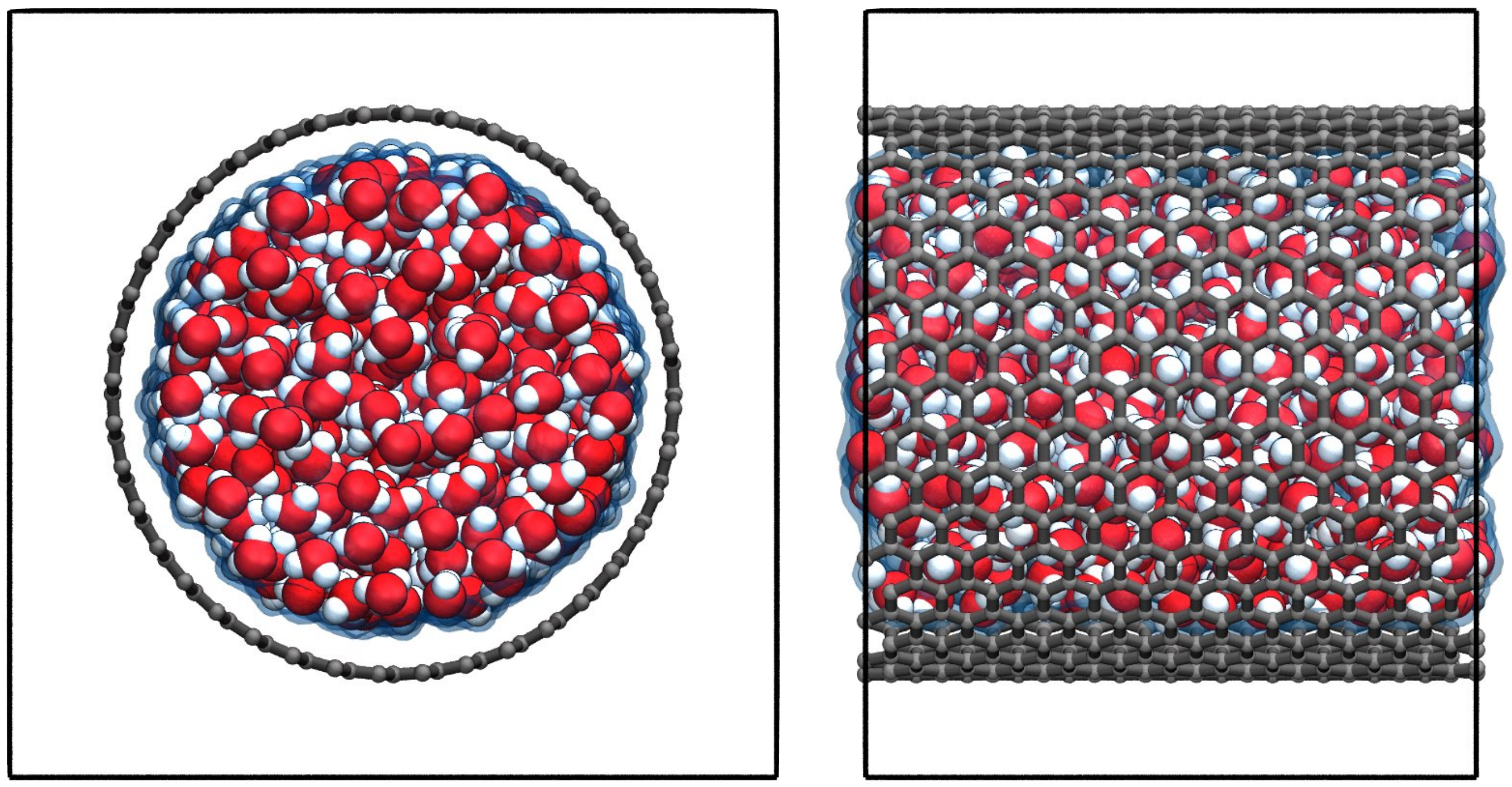}}\\
             & N$_{\mathrm{atoms}} = 2286$  & N$_{\mathrm{atoms}} = 2361$ & \\
             & N$_{\mathrm{water}} = 442$  & N$_{\mathrm{water}} = 467$ & \\
             &t$_{\mathrm{sim}} = 2$~ns & t$_{\mathrm{sim}} = 2.25$~ns & \\
            \cmidrule{1-4}
             \multirow{4}{*}{Nanotube (25,25)} &R$_\mathrm{tube} = 1.70$~nm
             & R$_\mathrm{tube} = 1.73$~nm& \multirow{4}{*}{\includegraphics[width=0.25\textwidth]{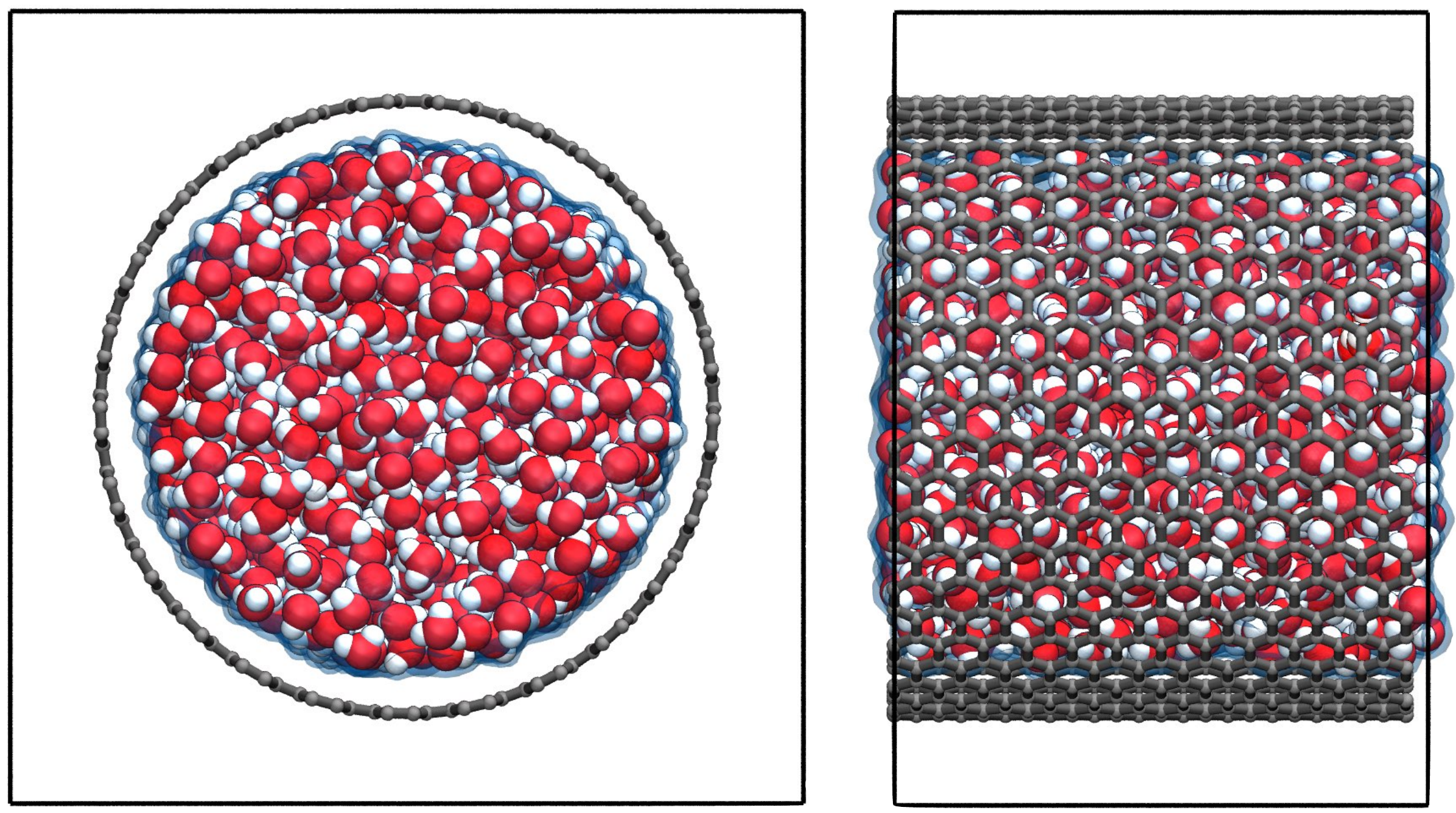}}\\
             & N$_{\mathrm{atoms}} = 3392$  & N$_{\mathrm{atoms}} = 3513$ & \\
             & N$_{\mathrm{water}} = 731$  & N$_{\mathrm{water}} = 771$ & \\
             &t$_{\mathrm{sim}} = 2$~ns & t$_{\mathrm{sim}} = 2$~ns & \\
            \cmidrule{1-4}
             \multirow{4}{*}{Nanotube (30,30)} &R$_\mathrm{tube} = 2.04$~nm
             & R$_\mathrm{tube} = 2.08$~nm& \multirow{4}{*}{\includegraphics[width=0.28\textwidth]{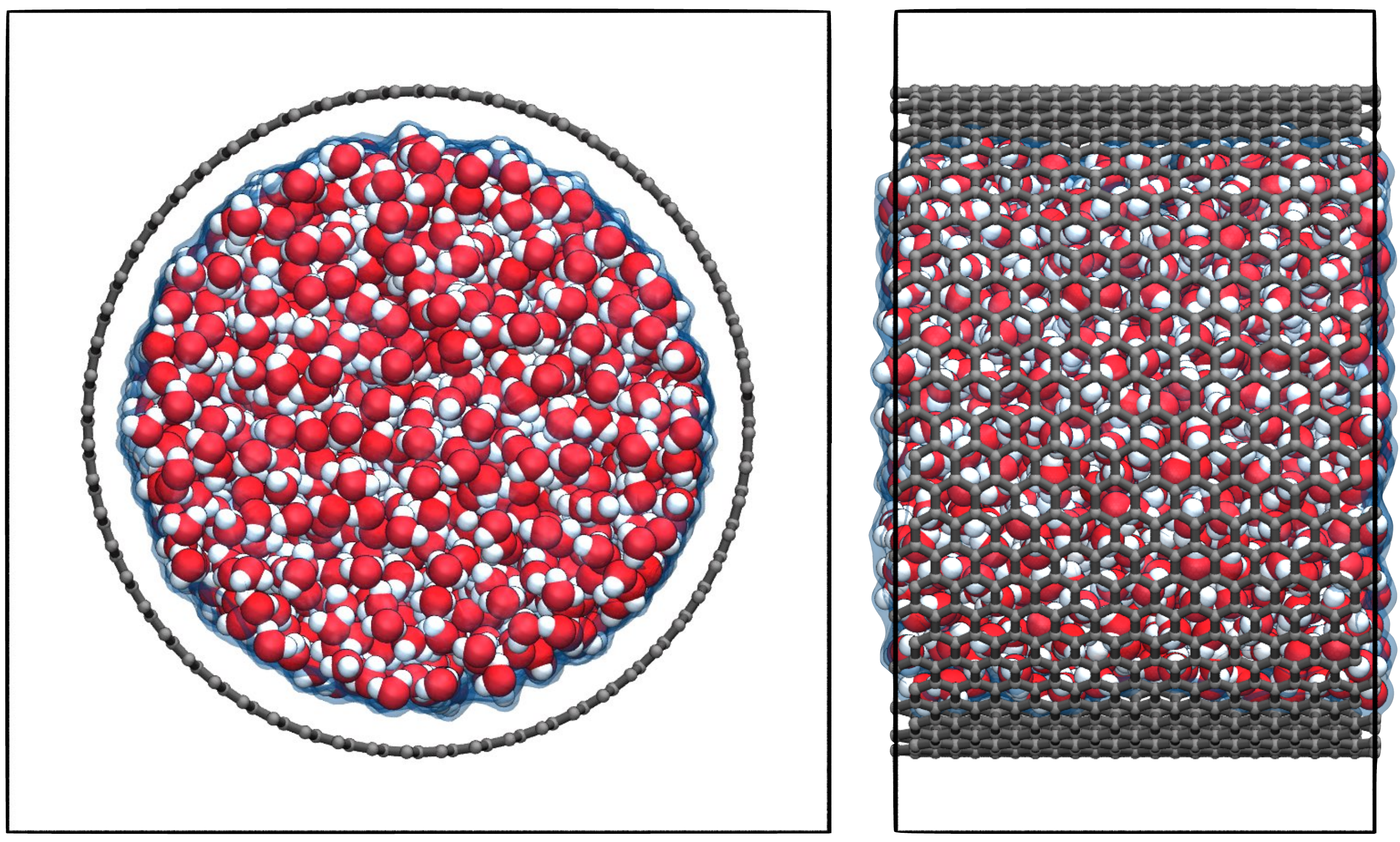}}\\
             & N$_{\mathrm{atoms}} = 4716$  & N$_{\mathrm{atoms}} = 4884$ & \\
             & N$_{\mathrm{water}} = 1092$  & N$_{\mathrm{water}} = 1148$ & \\
             &t$_{\mathrm{sim}} = 1.4$~ns & t$_{\mathrm{sim}} = 1.2$~ns & \\
            \cmidrule{1-4}
                         \multirow{4}{*}{Nanotube (35,35)} &R$_\mathrm{tube} = 2.38$~nm
             & R$_\mathrm{tube} = 2.42$~nm&\multirow{4}{*}{\includegraphics[width=0.29\textwidth]{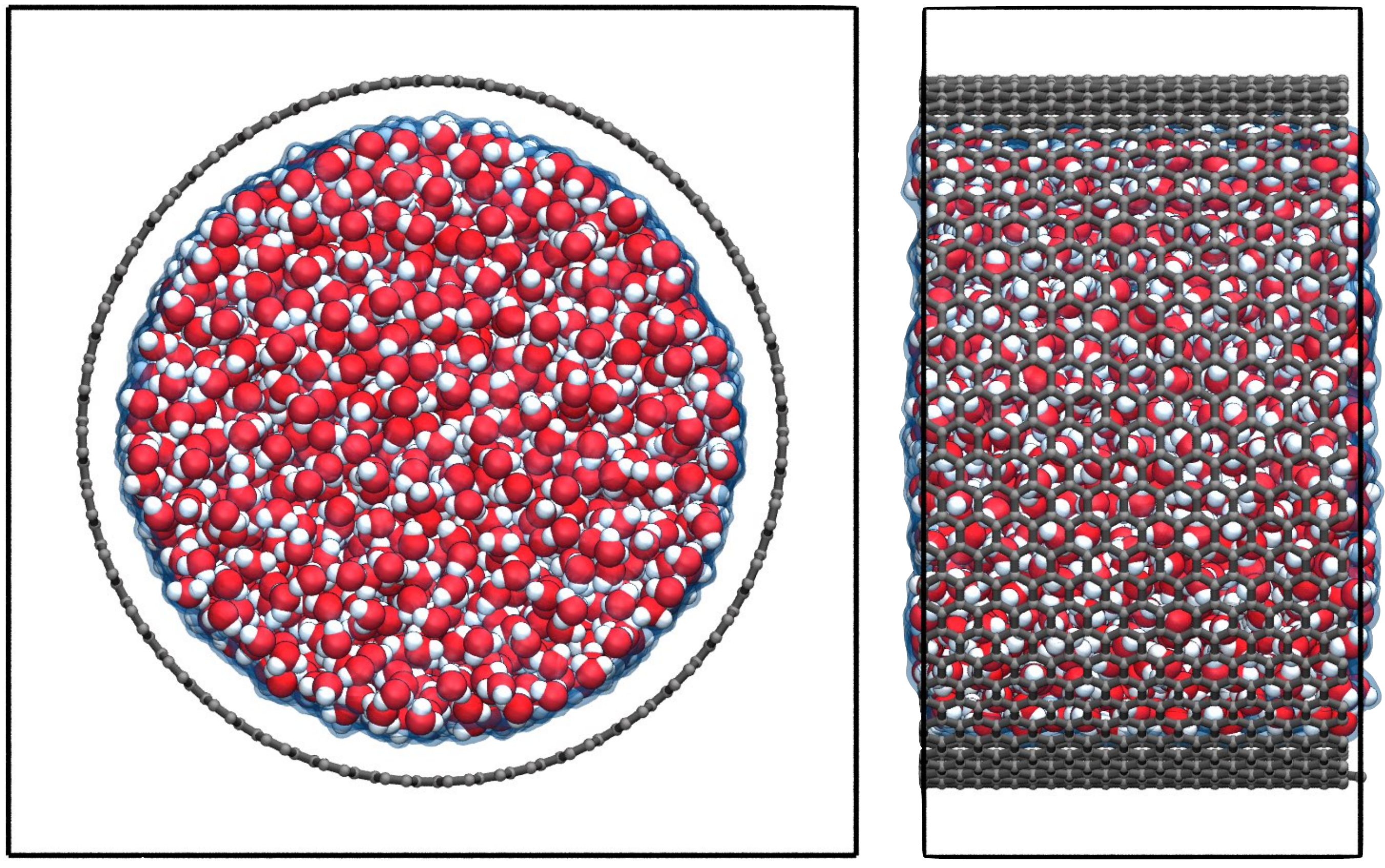}}\\
             & N$_{\mathrm{atoms}} = 6255$  & N$_{\mathrm{atoms}} = 6501$ & \\
             & N$_{\mathrm{water}} = 1525$  & N$_{\mathrm{water}} = 1607$ & \\
             &t$_{\mathrm{sim}} = 2$~ns & t$_{\mathrm{sim}} = 1$~ns & \\
             \cmidrule{1-4}
                                      \multirow{4}{*}{Nanotube (40,40)} &R$_\mathrm{tube} = 2.72$~nm
             & R$_\mathrm{tube} = 2.77$~nm&\multirow{4}{*}{\includegraphics[width=0.30\textwidth]{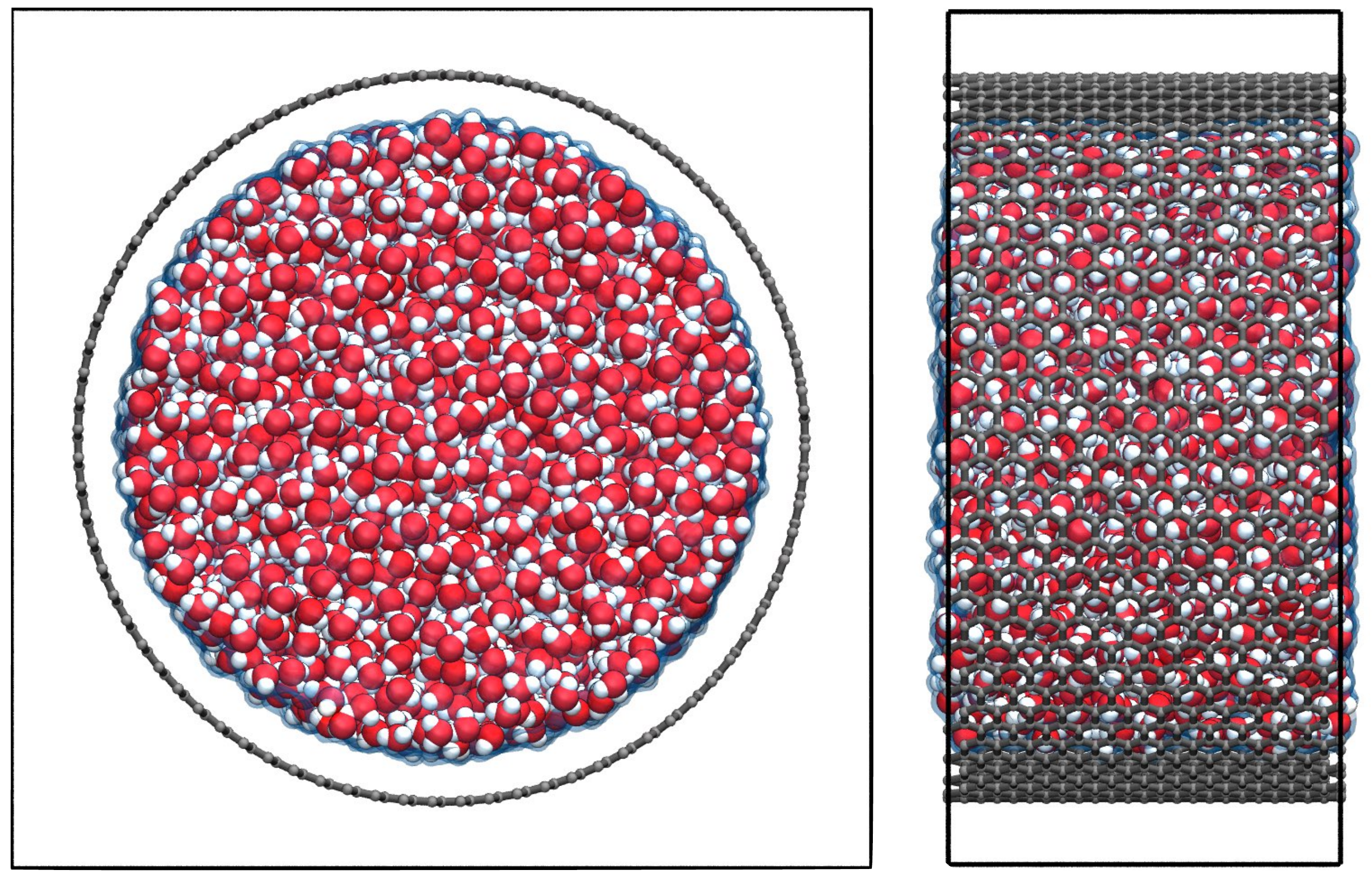}}\\
             & N$_{\mathrm{atoms}} = 8004$  & N$_{\mathrm{atoms}} = 8328$ & \\
             & N$_{\mathrm{water}} = 2028$  & N$_{\mathrm{water}} = 2136$ & \\
             &t$_{\mathrm{sim}} = 1.3$~ns & t$_{\mathrm{sim}} = 1$~ns & \\
             \cmidrule{1-4}
             \multirow{4}{*}{Sheet} &H$_\mathrm{film} = 3.5$~nm
             & H$_\mathrm{film} = 3.5$~nm&\multirow{4}{*}{\includegraphics[width=0.205\textwidth]{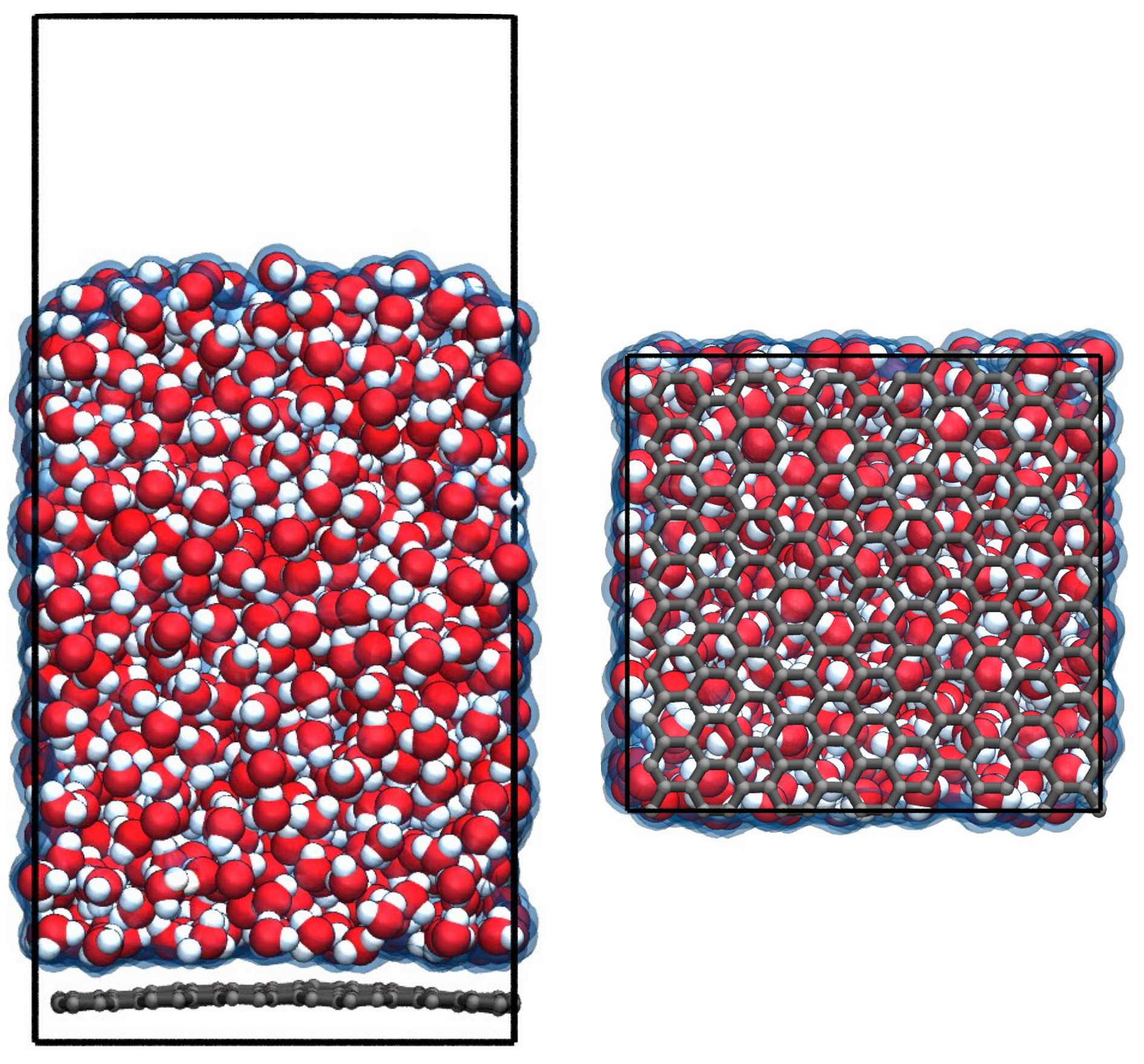}}\\
             & N$_{\mathrm{atoms}} = 2460$  & N$_{\mathrm{atoms}} = 2532$ & \\
             & N$_{\mathrm{water}} = 740$  & N$_{\mathrm{water}} = 764$ & \\
             &t$_{\mathrm{sim}} = 2.25$~ns & t$_{\mathrm{sim}} = 2$~ns & \\
             \cmidrule{1-4}
             \bottomrule
\caption{Detailed overview of systems investigated in this work. For each system, we report the total number of atoms, N$_\mathrm{atoms}$, the number of water molecules, N$_\mathrm{water}$, and the simulation length, t$_\mathrm{sim}$. For nanotubes and the flat sheet we further list the tube radii, R$_\mathrm{tube}$, and water film height, H$_\mathrm{film}$, respectively. In the right column, we show a front and side (bottom for the sheet) view of each system. The black lines represent the edges of the simulation box.} %
\label{tab:si_overview}
\end{longtable}

\subsubsection{Simulation setup}

For all molecular dynamics (MD) simulations, we employed the CP2K \cite{Kuhne2020} simulation package.
In this work, we performed three different types of MD simulations:
(i) \textit{ab initio} MD (AIMD) to generate training data for the development of our machine learning potentials (MLPs); (ii) MD simulations using classical nuclei employing MLPs to generate the results reported in the manuscript;
(iii) Path integral MD (PIMD) simulations with quantum nuclei to investigate the impact of nuclear quantum effects for the friction coefficient.  
In all three types of simulations, all atoms were treated as flexible and the center of mass velocity as well as the total angular momentum velocity were zeroed.
Moreover, deuterium masses was employed for the hydrogens to ensure a stable simulations at computationally feasible timesteps.

\subsubsection*{AIMD simulations}
The AIMD simulations of CNTs and BNNTs filled with water densities 0.6, 0.8, and 1.0 g/cm$^3$ were performed using Langevin MD at 330~K employing using a timestep of 1~fs.
We used the dual-space Goedecker-Tetter-Hutter pseudopotentials \cite{Goedecker1996} to represent the atomic cores and the Kohn-Sham orbitals were expanded in DVZP basis set of shorter range (SR) \cite{VandeVondele2007}.
In the original AIMD simulations a cutoff of 460 Ry was applied and the generalised gradient approximation (GGA) functional PBE \cite{Perdew1996B} with the D3 dispersion correction \cite{Grimme2010,Grimme2011} was used.
Within the training process of the MLPs, additional single point calculations were performed for the selected configurations where the cutoff was increased to 1050 Ry and the revPBE functional \cite{Zhang1998} with dispersion correction D3 was employed. %
The length of all AIMD simulations was above 100~ps.

\subsubsection*{MD simulations with classical nuclei}

Irrespective of the material and curvature, all simulations were performed in the NVT ensemble where the temperature was set to 300~K.
We employed two stochastic velocity rescaling thermostats \cite{Bussi2007} where one thermostat was applied to the liquid and solid, respectively.
The timestep was set to 1~fs.
To ensure that dynamical properties were not affected by the thermostats, the time constant was set to $1$~ps for both thermostats.
Each system was equilibrated for 50~ps before statistics were sampled for at least 1~ns.

\subsubsection*{PIMD simulations with quantum nuclei}

Path integral MD simulations in the NVT ensemble at 300~K were performed via thermostat ring polymer MD (TRPMD)~\cite{Rossi2014} employing the PILE thermostat~\cite{Ceriotti2010}.
Using deuterium instead of hydrogen allows us to achieve converged properties with 16 beads while employing a timestep 0f 0.25~fs.
Due the increased cost, simulations were performed for a reduced simulation time of 500~ps and using 6 instead of 12 replicas of the unit cell in axial direction.
To compute the friction coefficient, the summed force per frame was averaged over all beads which was then correlated to apply the established Green-Kubo relation.

\subsection{Development of machine learning potentials}
Our MLPs for water in CNTs and BNNTs rely on Behler-Parrinello neural network potentials (NNPs)~\cite{Behler2007,Behler2021}  in order to form a committee neural network potential (C-NNP)~\cite{Schran2020}.
The NNP formalism has been successfully employed to understand the unique properties of water~\cite{Morawietz2016,Cheng2019} and was also used to study water flow in hBN channels~\cite{Ghorbanfekr2020}, making it the ideal method for the current study.
Furthermore, we have recently shown that the C-NNP methodology enables the simple generation of MLPs for complex aqueous systems~\cite{Schran2021}.
The main idea behind the C-NNP approach is the combination of multiple NNPs in a ``committee model'', where the committee members are separately trained from independent random initial conditions to a subset of the total training set.
While the committee average provides more accurate predictions than the individual NNPs, the committee disagreement, defined as the standard deviation between the committee members, grants access to an estimate of the error of the model.
This committee disagreement provides an objective measure of the accuracy of the underlying model.
To construct a training set of such a model in an automated and data-driven way, new configurations with the highest disagreement can be added to the training set.
This is an active learning strategy called query by committee (QbC) and can be used to systematically improve a machine learning model.
For further details on the C-NNP methodology we refer the reader to Ref.~\citenum{Schran2020}.

\subsubsection{Automated development of the MLP}
For the development of two MLPs we used our established active learning workflow~\cite{Schran2020,Schran2021}, which is split into different generations.
In each generation new state points are targeted to yield a C-NNP that will be used to generate new candidate structures for the next generation.
This strategy relies on the fact that a C-NNP model can faithfully yield configurations even for thermodynamic conditions not yet considered in the training set, if the conditions are not drastically different from those already considered.
Within a generation, QbC is used to adaptively extend the training set by selecting the most representative configurations separately for each state point, improving its description.
With this procedure, multiple state points can easily be treated in parallel.
At the very beginning of each QbC cycle, a small number of random configurations is chosen to train the first committee, while in subsequent iterations, new configurations are selected based on the highest atomic force committee disagreement.
Convergence of these individual QbC cycles can be detected by monitoring this disagreement.
Once all QbC cycles for the selected conditions in a given generation are converged, the individual training sets are combined and a final tight optimization of that generation’s resulting NNP is performed.

\begin{figure*}[!ht]
    \centering
    \includegraphics[width=\textwidth]{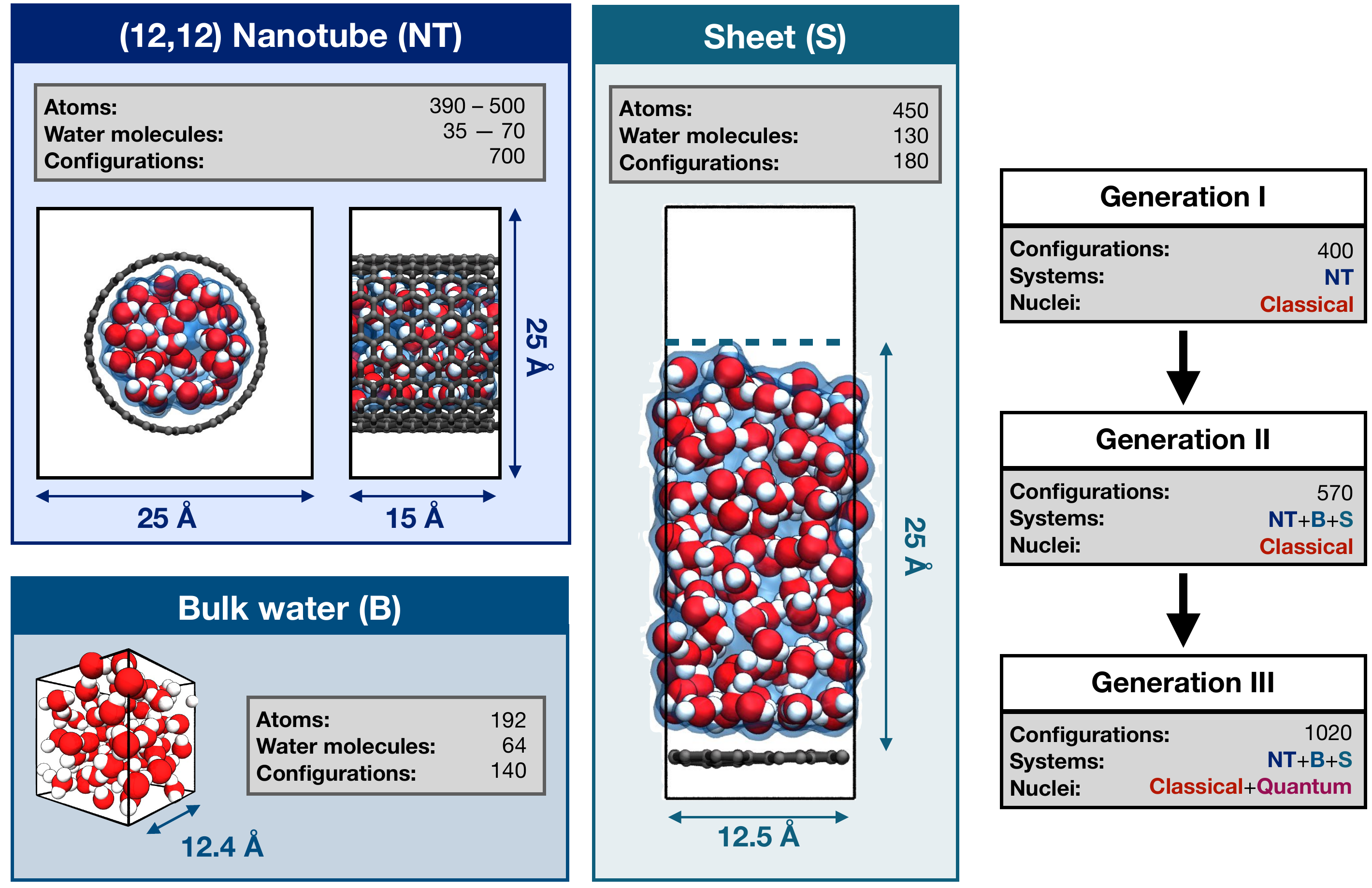}
    \caption{Overview of development of the MLP summarising the systems included in the stepwise active learning process.~\label{fig:model}
    The left part shows the three different types of systems incorporated in the model.
    These include a nanotube (NT) with chirality (12,12) filled with water at densities $\rho \in [0.6, 0.8, 1.0]$~g/cm$^3$, bulk water (B), and a water film on top of a sheet (S).
    For each of these systems, we show snapshots highlighting the relevant dimensions and giving information about the total number of atoms, the number of water molecules, as well as how many configurations of each system were included in the training of the final models.
    On the right, we show how the model is improved in a stepwise manner with our active learning workflow. 
    In Generation I, the model is exclusively build on configurations extracted from AIMD trajectories of water-filled nanotubes.
    Using this model as a computationally efficient structure generator, the training-set in generation II is extended by actively selecting individual structures from classical MD simulations performed with the model of generation I.
    In a final step, the model from generation II is employed to perform PIMD and configurations accounting for nuclear quantum effects (NQEs) are added to the training set of final model called generation III.
    For all generations, the configurations to be included in the training process where selected by QbC.}
\end{figure*}

Using this workflow as summarised in Fig.~\ref{fig:model}, we have first targeted small (12,12) carbon and hBN nanotubes filled with water densities of 0.6, 0.8 and 1.0 g/cm$^3$.
The initial models were seeded by short AIMD simulations at 330\,K, providing after active learning the first generation models with 465 and 387 structures for carbon-water and hBN-water, respectively.
In the second generation, bulk water and water on flat graphene and hBN sheets were targeted by individual active learning cycles.
The required structures were generated from C-NNP simulations at 330\,K with a minimum length of 200\,ps for both materials.
Bulk water was simulated in a periodic box of 64 water molecules at experimental equilibrium density, while water in contact with the flat sheets was described by 130 and 136 water molecules on a 3x5 supercell of graphene and hBN, respectively.
Overall, for both materials 148 new structures from these conditions were added to the training sets of the two models.
To incorporate the quantum nature of the nuclei into the description of our C-NNP models, we performed PIMD simulations in the next generation.
For that purpose, we performed the same simulations as before with quantum nuclei using the two C-NNP models, resulting in 5 distinct conditions which have been targeted by separate active learning cycles.
During these cycles 483 and 350 additional structures were identified and added to the training set.

\subsubsection{Validation of the MLP}

The final training sets consist of 1096 and 885 structures for CNNTs and BNNTs, respectively, while spanning diverse conditions and including both a classical and quantum description of the nuclei as shown in figure~\ref{fig:model}.

\begin{figure*}[ht]
    \centering
    \includegraphics[width=\textwidth]{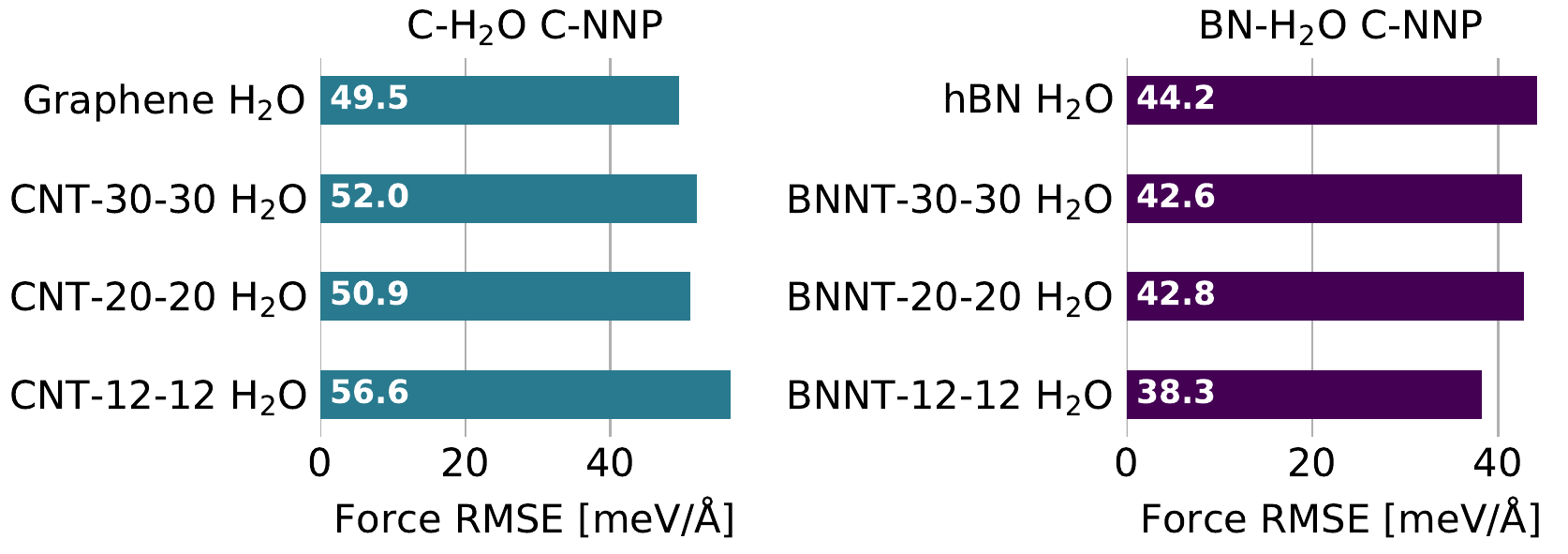}
    \caption{Performance of the carbon-water and hBN-water C-NNP models for the diverse validation sets.
    The left panel summarises the force accuracy of carbon-water C-NNP model, while the right panel features the accuracy of the hBN-water model.
    For each model a diverse validation set was constructed from our production simulations for nanotubes of various size as well as water in contact with the flat sheets.
    \label{fig:validation}.
    }
\end{figure*}

For our carbon-water C-NNP model we obtain an energy and force root mean square error (RMSE) for the training set of 0.8\,meV per atom and 72\,meV/\AA{}, respectively.
The hBN-water C-NNP model was trained with an energy and force RMSE value of 0.7\,meV per atom and 47\,meV/\AA{}, respectively.
To test the performance of the model for the study of water flow in single-wall nanotubes of different radii, we build an out-of-sample validation set by drawing configurations from our production simulations for the 12-12, 20-20 and 30-30 nanotubes as well as water on the flat graphene and hBN surfaces.
Evaluating the DFT energies and forces for 50 randomly selected structures of each system allows us to assess the accuracy of our models in application to faithfully validate the main findings of this study.
The performance of our two models over this diverse set of validation structures is summarised in figure~\ref{fig:validation}.
The largest force RMSE for this validation set is 63 mev/\AA{}, suggesting that the model remains accurate and robust across the wide range of conditions and systems studied here.

\begin{figure*}[ht]
    \centering
    \includegraphics[width=0.9\textwidth]{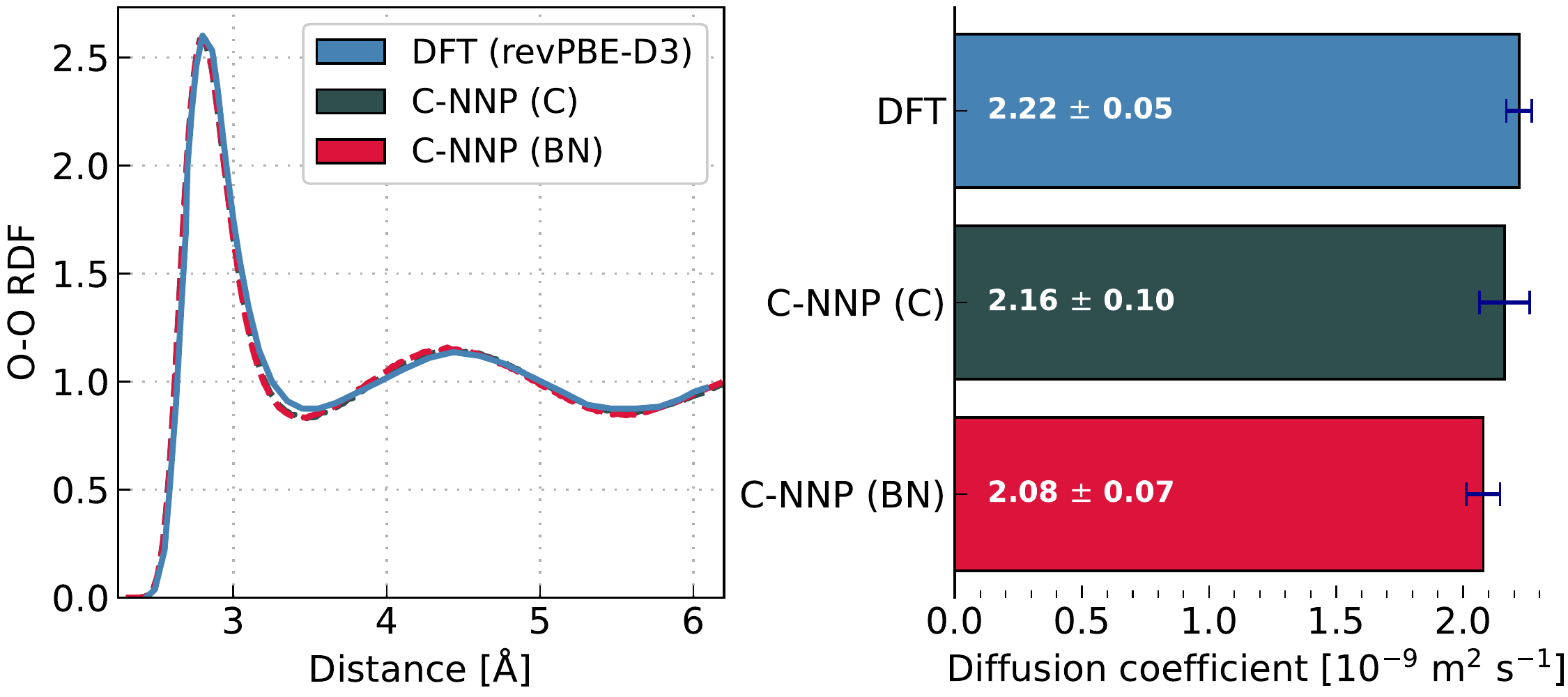}
    \caption{Properties of bulk water predicted by the published simulations \cite{Marsalek2017} performed with the chosen DFT functional and by our C-NNPs.
    The left panel shows the oxygen radial distribution function (RDF) and on the right we illustrate the performance for the diffusion coefficient.
    The error bars correspond to the statistical error obtained from splitting the entire trajectory into two blocks.
    \label{fig:bulk}.
    }
\end{figure*}

To further validate the performance of our model for the description of water, we ran bulk water simulations at ambient conditions and compare to published \cite{Marsalek2017} structural and dynamical properties obtained with our reference DFT setup as shown in Fig.\ref{fig:bulk}.
To achieve a fair comparison, hydrogen masses were adopted instead of deuterium and we timestep was decreased to 0.5 fs.
For both models, the simulation time was 1~ns.
The radial distribution functions (RDF) obtained with both our carbon-water and hBN-water model are in near perfect agreement with the published results of the revPBE-D3 functional \cite{Zhang1998,Grimme2010,Grimme2011}.
Furthermore, following the procedure outlined in reference \citenum{Marsalek2017} we computed the diffusion constants of water based on a linear fit to the mean square displacement of the molecular center of mass in the time interval between 1 and 10~ps.
The correction introduced by Yeh and Hummer was applied to account for finite size effects \citenum{Yeh2004}.
Using these settings, we find that both models are in very good agreement with the reference, accentuating the high accuracy of our model.

\subsection{Computation of properties}
\subsubsection{Friction coefficient}
The friction coefficient $\lambda$ is defined as the ratio between the total force of the solid exerted on the liquid parallel to the surface, $\mathbf{F_\mathrm{wall}}$, normalised by the surface area, $A$, and the fluid velocity at the interface, $v_\mathrm{wall}$,
\begin{equation}
    \lambda= \frac{\mathbf{F_\mathrm{wall}}}{A v_\mathrm{wall}} \mbox{ . }
\end{equation}
Based on linear response theory, the friction coefficient can be readily computed from equilibrium MD simulations using a Green-Kubo relation \cite{Bocquet1994,Bocquet2013}
\begin{equation}
    \lambda_{\mathrm{GK}} (t)= \frac{1}{A k_\mathrm{B}T} \int_0^t \langle \mathbf{F_\mathrm{wall}} (t') \cdot \mathbf{F_\mathrm{wall}} (0)\rangle dt' \mbox{ , }
    \label{eq:friction_GK}
\end{equation}
where $k_\mathrm{B}$ is the Boltzmann constant, $T$ corresponds to the temperature, and the brackets correspond to an ensemble average.
The total force of the solid, $\mathbf{F_\mathrm{wall}}$, given by the summed force of all solid atoms of a given configuration, is saved every timestep (1 fs).
For nanotubes, only the force components in axial ($z$) direction were used to compute the friction.
In the case of graphene and hBN, conversely, $\mathbf{F_\mathrm{wall}}$ was evaluated for both in-plane dimensions ($x,y$).
The friction coefficient of the system is then recovered for sufficiently long time intervals, $\lambda = \lim_{t \to \infty} \lambda_\mathrm{GK}$.
In practice, however, rather than reaching a plateau at infinite time the Green-Kubo integrals vanish for systems of finite size \cite{Bocquet1997} complicating its accurate evaluation.
In this work, we circumvent this so-called ''plateau problem'' \cite{Espanol2019} by following a recently introduced methodology \cite{Oga2019} where the Green-Kubo integrals obtained from simulations are fitted to an analytical expression.
The finite-size-corrected friction coefficient can then be computed based on the fitting parameters.
For all systems, the Green-Kubo integrals were computed for a correlation time of 2.5~ps while an $R^2$ value above 0.85 for the fit of the analytical expression from reference \cite{Oga2019} was ensured.
\\

\subsubsection{Free energy surface}

In the manuscript, we link the friction coefficient to the corrugation free energy surface (FES) of the oxygen and hydrogen atoms in the contact layer.
Here, we provide a detailed explanation of the methodology behind the computation of these profiles.
Following previous work \cite{Tocci2014a,Tocci2020}, the two-dimensional FES is given by $\Delta F (x,y) = - k_B T \ln p_i (x,y)$, where $k_B$ is the Boltzmann constant, $T$ represents the temperature (here 300~K), and $p_i (x,y)$ corresponds to the normalised two-dimensional probability of finding species $i \in [\mathrm{O},\mathrm{H}]$ in the contact layer at a point $(x,y)$.
With nanotubes being of one-dimensional character, the coordinates $x$ and $y$ correspond to the azimutal and axial direction.
Based on these definitions, computing the FES and related properties requires (i) a consistent definition of the contact layer, (ii) the efficient sampling of $p_i (x,y)$ in cylindrical-like systems, and (iii) an appropriate measure quantifying the corrugation itself.

Starting with the contact layer, we computed the distribution of the distance between the oxygen of each water molecule $j$ and the closest solid atoms (C, B, N), $d_{\mathrm{O}-\mathrm{X}}^j$.
Interestingly, we found that for all systems the first minimum was located at a distance of $\approx 4.8 \mathrm{\AA}$.
Therefore, we assign a water molecule $j$ to the contact layer if $d_{\mathrm{O}-\mathrm{X}}^j \leq 4.8 \mathrm{\AA}$.
Going through the configurations of a trajectory, the positions of either the oxygens or hydrogens - dependent on the species for which the FES is computed - of water molecules satisfying the criterion above are selected.
By decomposing these coordinates into an axial and angular contribution, these positions are projected onto a two-dimensional grid corresponding to the nanotube opened up along a cut in axial direction.
At this point, we already have an estimate of $p_i (x,y)$, however, we can accelerate its convergence by mapping $p_i (x,y)$ of the entire tube onto its unit cell.
Making use of the inherent periodicity of the crystal lattice of the nanotube allows to extract 30-480 times more statistics per configuration dependening on the system. 
Following this procedure and by choosing a high sampling frequency of 20\,fs, we obtain a converged estimate of $p_i (x,y)$ and, thus, the FES.
At last, we are only left with defining the corrugation of the FES.
In analogy to previous work \cite{Tocci2014a,Tocci2020}, this is done by taking the highest free energy present in the FES.
However, to ensure this quantity is independent of noise we smooth the FES in advance by applying a two-dimensional Savitzky-Golay filter as implemented in common python libraries such as SciPy \cite{Virtanen2020}.

\section{Sensitivity of the friction coefficient}
To ensure that the main results reported in the manuscript are invariant with respect to simulation settings, in this section we perform a sensitivity analysis of the friction coefficient.
If not stated otherwise, the simulation time for each system is at least 1~ns.\\

\subsection{Water density}
\begin{figure*}[h]
    \centering
    \includegraphics[width=\textwidth]{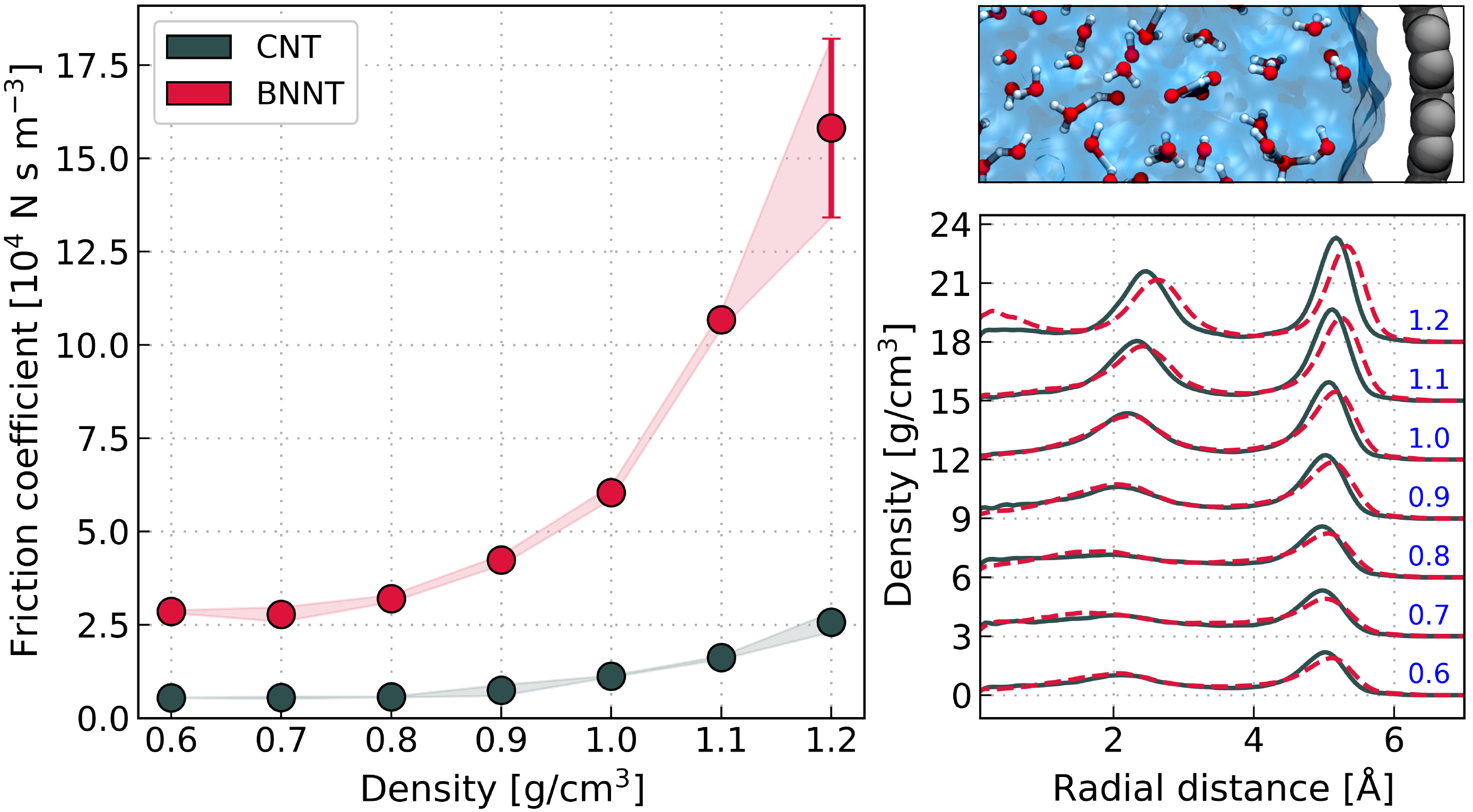}
    \caption{ Impact of the water density on the friction coefficient. 
    The left panel shows the friction coefficient computed for the smallest nanotubes (12,12) as function of the density.
    The error bars correspond to the statistical error obtained from splitting the entire trajectory into two blocks.
    In the right panel we show at the bottom the density profiles for the different densities analysed as a function of the radial distance from the tube center of mass.
    For the sake of clarity, the profile are shift by a factor of 3 along the y-axis with increasing density.
    The continuous lines correspond to CNTs while BNNTs are plotted using dashed lines.
    The top panel shows a snapshot of a CNT to illustrate the radial dependence more clearly.
    \label{fig:friction_density}}
\end{figure*}
To investigate the impact of the water density on the friction coefficient we performed additional simulations on the smallest nanotubes (12,12) varying the water density.
A summary of this analysis is shown in the left panel of figure \ref{fig:friction_density}.
We observe that for both materials the friction remains constant for a density $\rho \leq 0.8$ g/cm$^3$ resulting in values of $\approx 0.5 \cdot 10^4$~N~s~m$^{-3}$ and $\approx 3.0 \cdot 10^4$~N~s~m$^{-3}$ for the CNT and BNNT, respectively. 
At larger densities, conversely, the friction coefficient increases significantly reaching a five-fold enhancement at the highest density of $1.2$~g/cm$^3$.
Comparing the mass density profiles as a function of the radial distance illustrated in the right panel of figure \ref{fig:friction_density} gives insight into the origin of the enhancement of the friction.
In case of densities $\rho \leq 0.8$ g/cm$^3$, the water is in the liquid state.
When the number of water molecules inside the tubes increases, however, the fluid becomes more  structured (see the curves for $\rho = 0.9$ g/cm$^3$) and eventually freezes.
This applies in particular to densities $\rho \geq 1.0$ g/cm$^3$ where a second peak close to the tube center becomes apparent indicating the formation of a double-ring structure.
The observation that a phase transition to the solid state induces a significantly larger friction is of general interest and could spark future research.
With respect to this work, however, we lack knowledge of the physical water density inside the tubes of varying diameters at equilibrium.
In principle, the accurate number of water molecules in each tube could be determined by performing either Grand canonical Monte Carlo (GCMC) simulations or connecting the nanotube to two reservoirs as done in previous work \cite{Falk2010,Falk2012}.
Rather than performing these expensive and from the MLP perspective challenging simulations, here we employ the approximation of setting the density in all nanotubes to the bulk limit of $1.0$~g/cm$^3$.
While this approach might seem a bit crude, we note that (i) the difference between the materials hardly changes throughout all densities studied and (ii) the error associated with this approximation decreases with tube radius.
Therefore, we expect that the main findings presented in the manuscript still hold despite the change of the absolute values.

\subsection{System size}
\label{sec:si_size}
\begin{figure*}[h]
    \centering
    \includegraphics[width=\textwidth]{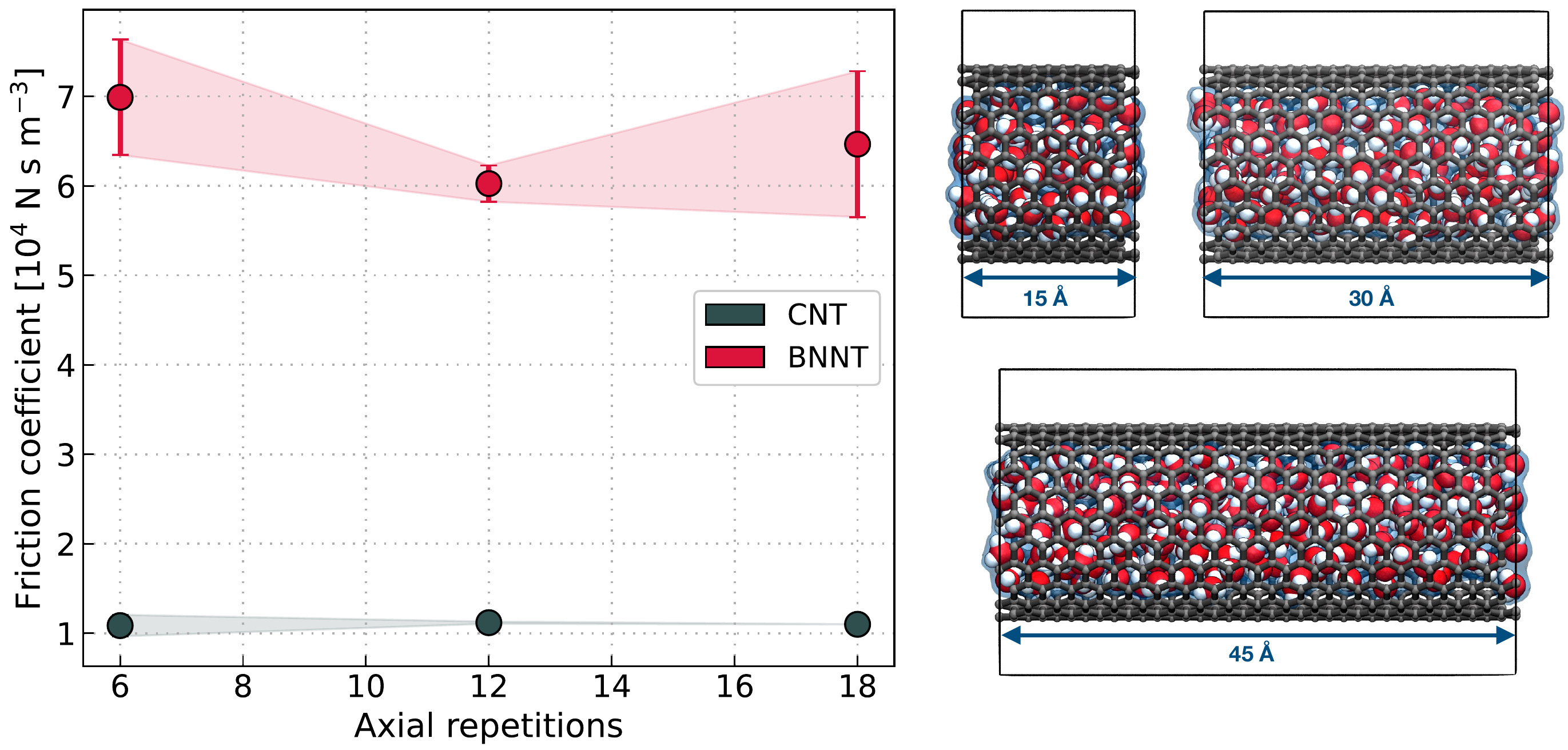}
    \caption{Dependence of the friction coefficient on the system size.
    The system size is quantified by the number of repetitions of the unit cell of a nanotube in axial direction corresponding to $\approx 2.5$~$\mathrm{\AA}$.
    The three systems shown here, correspond to an axial length of $\approx 15$, 30, and 45~$\mathrm{\AA}$ comprising roughly 500, 1000, and 1500 atoms, respectively.
    The error bars correspond to the statistical error obtained from splitting the entire trajectory into two blocks.
    On the right, we show snapshots of the different tubes where arrows illustrate the axial length.
    \label{fig:friction_size}}
\end{figure*}
While the usage of MLPs allows to reach significantly longer time and length scales, the simulation of the large diameter nanotubes are still computationally demanding.
Therefore, for all nanotubes the box length in axial direction ($z$) is set to repetitions of the unit cell of the nanotube corresponding to $\approx 30$~$\mathrm{\AA}$.
In figure \ref{fig:friction_size}, we show based on the smallest nanotubes (12,12) that this size is indeed sufficient to obtain converged results for the friction coefficient.
For both materials, the additionally simulations on cells comprising 6 and 18 axial repetitions, respectively, yield almost identical results for the friction of water inside the nanotubes.
With results remaining unchanged even for an axial dimension smaller than the tube diameter, we are confident that the employed system size is sufficient to obtain converged friction coefficients even for the largest nanotubes.

\subsection{Simulation time}
\begin{figure*}[t]
    \centering
    \includegraphics[width=0.5\textwidth]{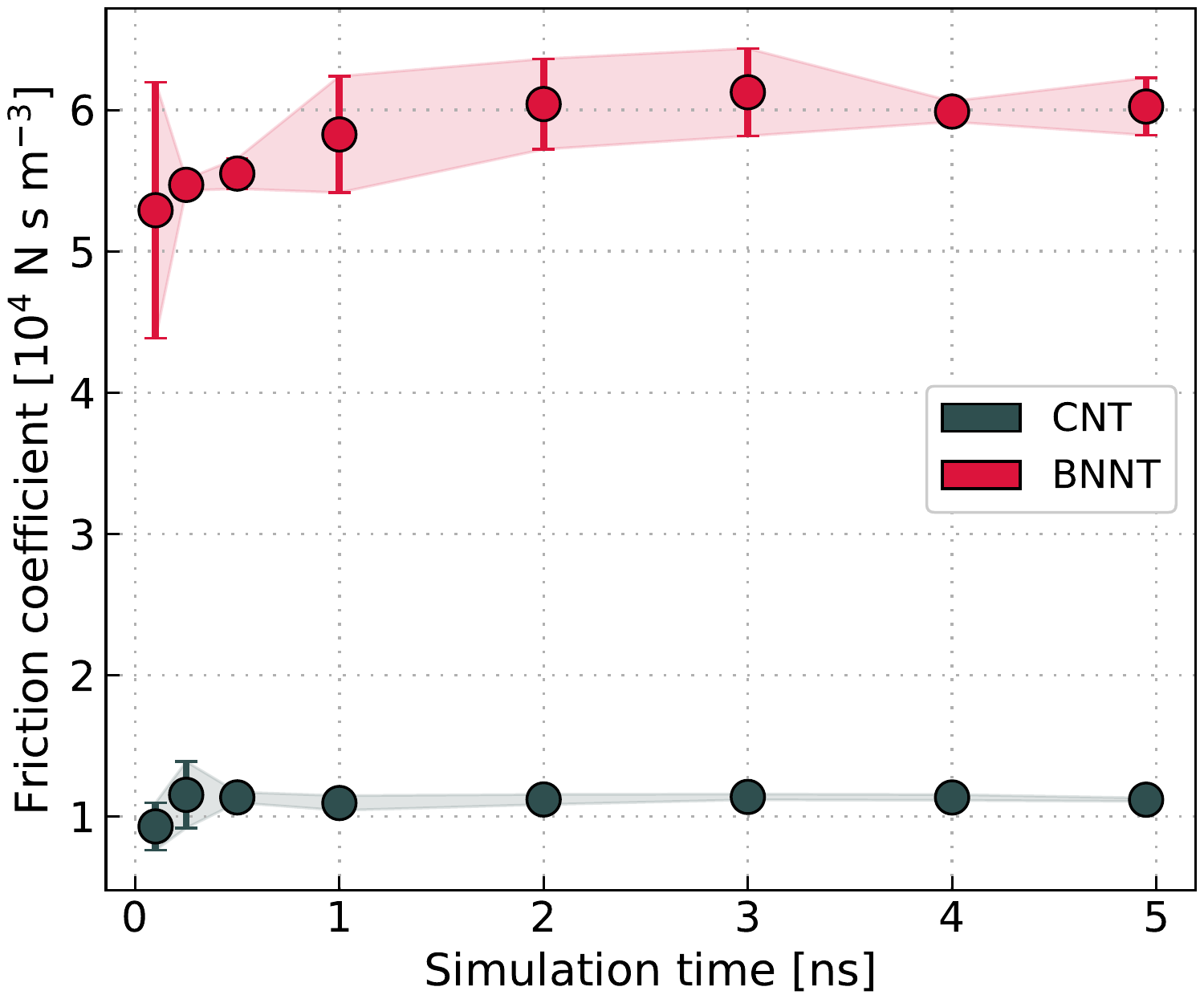}
    \caption{Convergence of the friction coefficient with simulation time.
    The error bars correspond to the statistical error obtained from splitting the trajectory of the given length into two blocks.
    \label{fig:friction_time}}
\end{figure*}
In analogy to analysing the impact of the system size, here we explore the convergence of the friction coefficient with respect to the simulation length.
To this end, we computed the friction coefficient for the smallest nanotubes (12,12) by varying the production time between 100~ps and 5~ns as shown in figure \ref{fig:friction_time}.
Similar to the results presented in the previous section, the friction coefficient is found to change relatively little with increasing simulation time.
For both materials, trajectories of the short length of 100~ps are already sufficient to converge the friction within $\approx 85\%$ with respect to the value obtained from the reference simulation of 5~ns.
While these short simulations have relatively large statistical errors, this convergence check shows that previous AIMD simulations of comparable simulation length \cite{Tocci2014a,Tocci2020} can indeed provide reliable results within the given uncertainty.
For our work, conversely, where all systems have been simulated for at least 1~ns this analysis confirms that the estimates of the friction coefficients are reliable and converged with respect to the length of the trajectory.

\subsection{DFT functional}
\label{sec:functional}
\begin{figure*}[t]
    \centering
    \includegraphics[width=\textwidth]{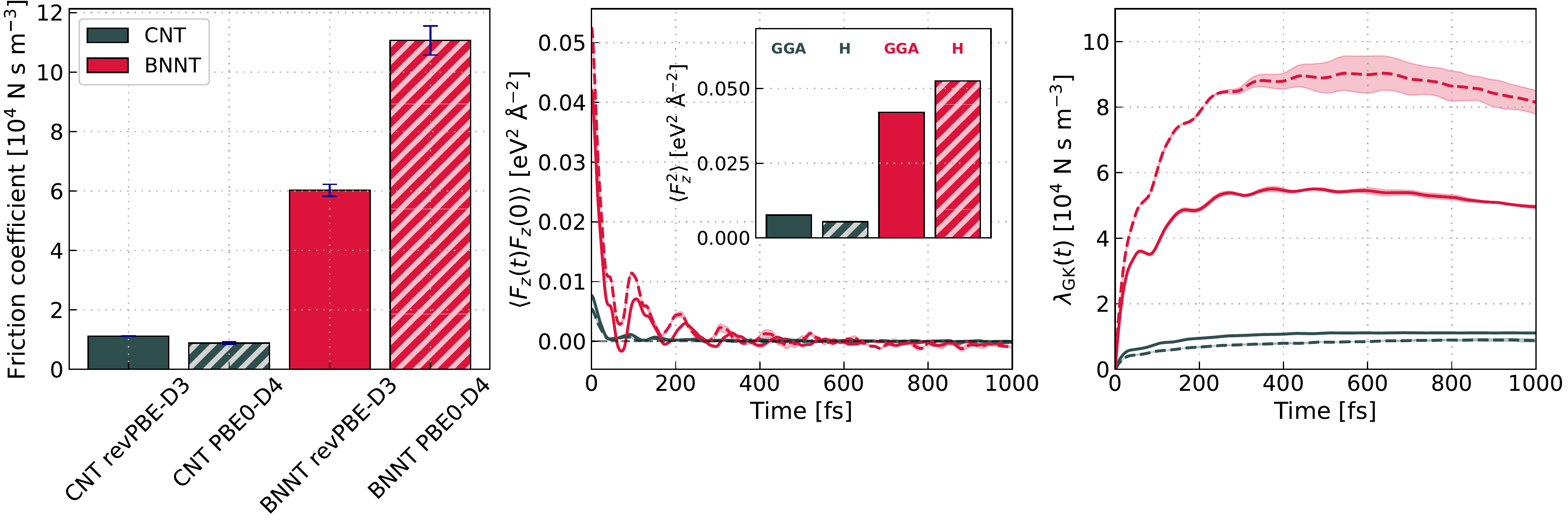}
    \caption{Dependence of the friction coefficient on the DFT functional.
    On the left the extracted friction coefficients based on the approach presented in reference \cite{Oga2019} are shown.
    The center panel shows the autocorrelation function of the summed force in axial-direction (z) as a function of time for the smallest CNT and BNNT.
    Dependent on the level of theory of the simulation, the lines are either solid or dashed corresponding to GGA and hybrid (H) DFT.
    The inset shows a bar plot of the ensemble average of the squared summed force.
    In the right panel, we show the estabilshed Green-Kubo relation where friction coefficient is proportional to integrated autocorrelation function.
    The errorbars and shaded areas around the curves correspond to the statistical error obtained by splitting the trajectory into two blocks.
    \label{fig:friction_functional}}
\end{figure*}

In this subsection, we investigate how the friction coefficient of water changes with the chosen DFT functional.
To this end, we took the training set of our MLP and recomputed energies and forces based on the hybrid functional PBE0 \cite{Perdew1996A,Adamo1999} and the improved dispersion correction D4 \cite{Caldeweyher2017,Caldeweyher2019,Caldeweyher2020}. 
It is important to note that, similar to its GGA analogue PBE \cite{Gillan2016}, we found that PBE0-D4 predicts a significant overstructured liquid water in the bulk phase.
This is distinctly different from the structure obtained with revPBE-D3 which is in close agreement with experiments \cite{Marsalek2017}.
Based on this functional and the associated new dataset, we trained new MLPs for the two materials investigated in this study.
To analyse the dependence of the friction coefficient on the chosen level of theory, we then perform additional simulations on the smallest nanotubes (12,12), compute the friction coefficients, and compare them to those obtained with our models based on revPBE-D3.
A summary of this comparison is shown in the left panel of figure \ref{fig:friction_functional}.
Interestingly, the hybrid DFT level predicts a similar friction coefficient of water inside the CNT while the values for the BNNT deviate almost by a factor of two with the hybrid predicting the higher friction.
To understand the origin behind these trends we take a closer look at the autocorrelation function of the summed force in axial (z) direction shown in the center panel of figure \ref{fig:friction_functional}.
While the curves look almost identical for the CNT, for the BNNT we observe that the autocorrelation function is shifted by a constant value while the general trends and oscillations are very similar.
This shift can be traced back to the larger average squared summed force $\langle F_z^2 \rangle$ for the hybrid in case of BNNTs.
This is also supported by the friction coefficient following the established Green-Kubo relation \cite{Bocquet1994,Bocquet2013} based on the integrated autocorrelation shown in the right panel of figure \ref{fig:friction_functional}.
A reason for this difference in the root mean square force might stem from the generally observed increase of the bandgap going from a GGA to a hybrid functional.
This way, the electrostatic interactions between the nitrogen and hydrogen are enhanced binding the water molecule stronger to the surface during the docking events and, thus, leading to a larger friction.

\subsection{Nuclear quantum effects}
\begin{figure*}[t]
    \centering
    \includegraphics[width=0.5\textwidth]{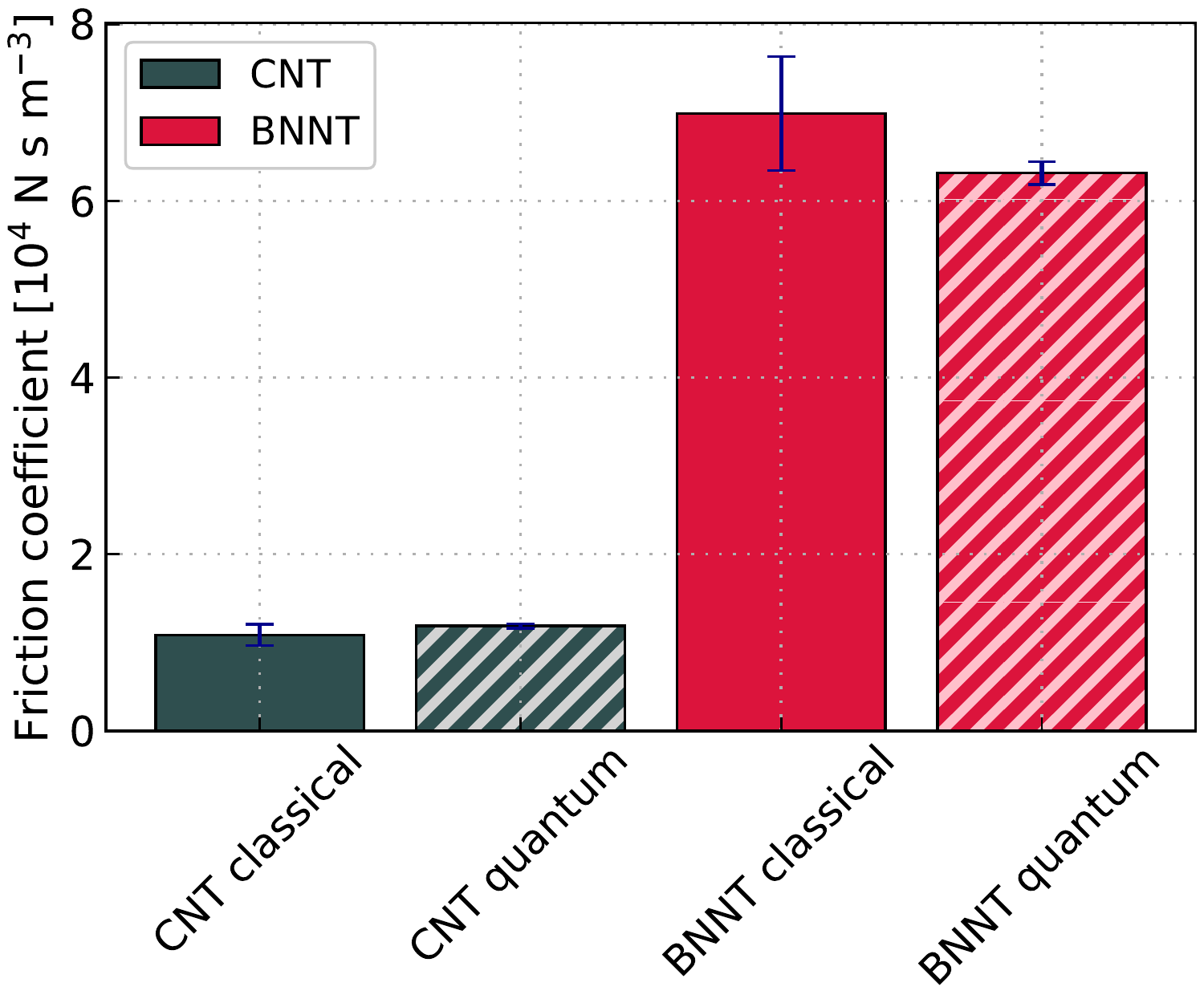} \caption{Impact of nuclear quantum effects on the friction coefficient.
    The friction coefficient were extracted from the established Green-Kubo relation using the approach presented in reference \cite{Oga2019}.
    The errorbars and shaded areas around the curves correspond to the statistical error obtained by splitting the trajectory into two blocks.
    \label{fig:friction_nqes}}
\end{figure*}

Due to the low mass of the proton's in water it is worth investigating the impact of nuclear quantum effects (NQEs) on the friction coefficient.
In figure \ref{fig:friction_nqes} we compare the values for the smallest nanotubes (12,12) computed with MD with classical nuclei and T-RPMD which includes the quantum nature of the nuclei.
Due to the high computational cost associated with PIMD, the number of axial repetitions was reduced to 6 instead of 12 for all simulations to enable a fair comparison.
In section \ref{sec:si_size} we showed that for the smallest tubes the numerical values change only to a small extend.
Overall, NQEs have an almost negligible impact on the friction coefficient with the MD and T-RPMD estimates agreeing with each other within error bars.

\subsection{Tube flexibility}

 It is well known that the coupling between the phonon modes of the confining material and the water vibrations strongly affects the fluid transport across low-dimensional materials \cite{Ma2015,Marbach2018}.
 Treating the confining solid as flexible and allowing the atoms to fluctuate has, thus, a crucial impact on the predicted water flux. 
 A previous study based on classical force fields has indeed shown that water experiences a significantly lower friction in flexible compared to rigid CNTs \cite{Sam2017}.
 To get a ``fair'' comparison to simulation studies treating the nanotubes \cite{Falk2010} or graphene and hBN sheet \cite{Poggioli2021} as rigid, here we compute the friction coefficient for rigid nanotubes.
 Similar to the other convergence checks, we limit this analysis to the smallest CNT and BNNT (12,12).
 When simulating the nanotubes as rigid bodies, we ensured that the error estimate provided by the committee disagreement in the C-NNP formalism remains sufficiently small underlying that the MLP provides a reliable estimate of the friction coefficient.

\begin{figure*}[t]
    \centering
    \includegraphics[width=\textwidth]{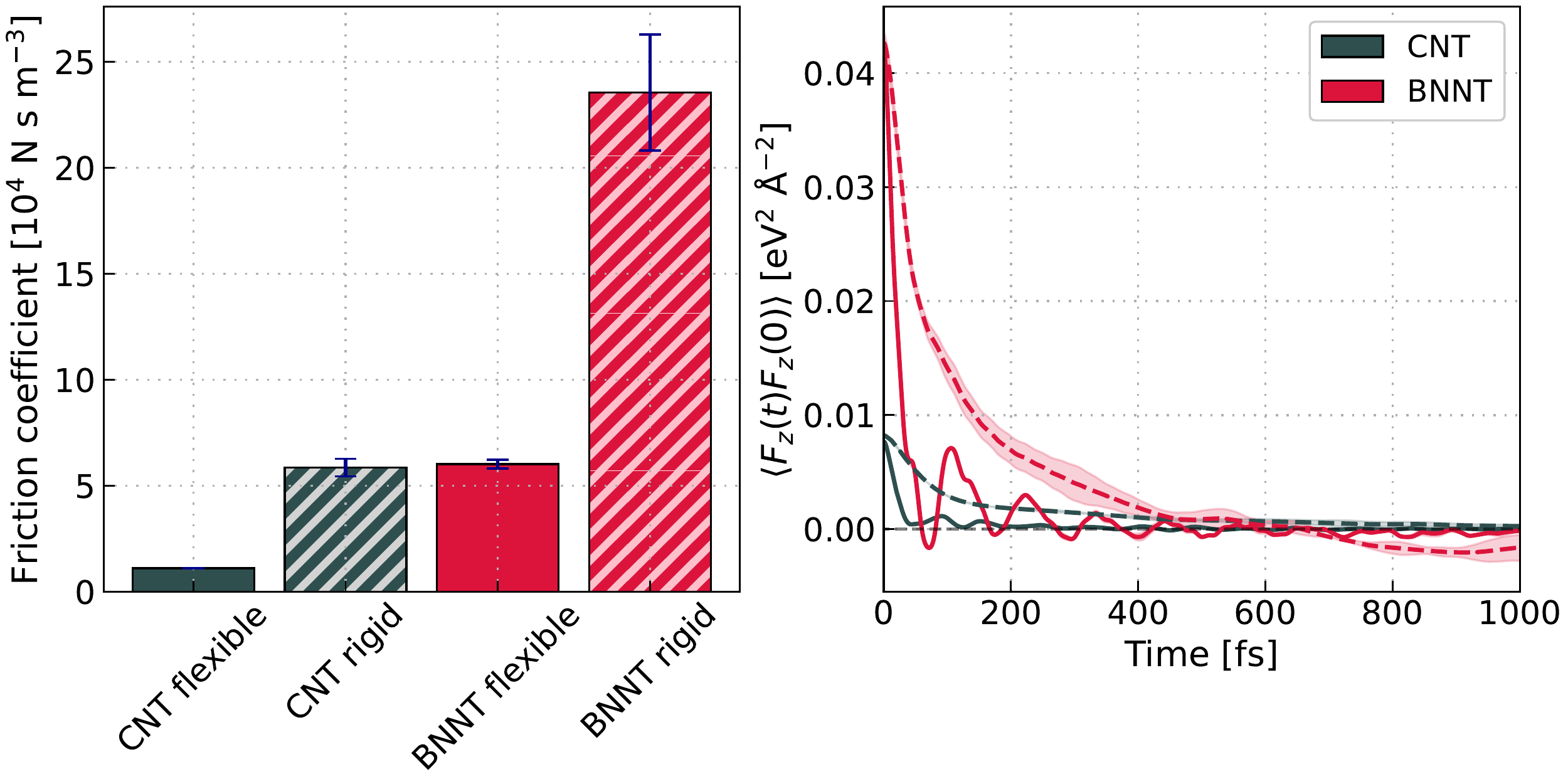}
    \caption{Impact of the flexibility of the tube walls on the friction coefficient.
    On the left the extracted friction coefficients based on the approach presented in reference \cite{Oga2019} are shown.
    In the right panel we show the autocorrelation function of the summed force in axial-direction (z) as a function of time for the smallest CNT and BNNT.
    The solid lines correspond to treating the walls as flexible and represent the results presented in the manuscript.
    Dashed lines, conversely, illustrate the autocorrelation function in rigid nanotubes.
    The error bars on the left and shaded areas on the right correspond to the statistical error obtained by splitting the trajectory into two blocks.
    \label{fig:friction_flexible}}
\end{figure*}

 The summary of this analysis is shown in the left panel of figure \ref{fig:friction_flexible}.
 In line with previous simulations \citenum{Sam2017} and theory \cite{Marbach2018}, we observe for both materials a $\approx 4 - 5$ times larger friction when the solid is treated as rigid.
 We also compare the summed force autocorrelation function in the left panel of \ref{fig:friction_flexible}.
 In contrast to our findings for the functional dependence in section \ref{sec:functional}, the static contribution to the friction, $\langle F_z^2 \rangle$, is almost independent of the flexibility of the tube.
 Conversely, the oscillations of the autocorrelation function vary greatly indicating that the coupling between fluid and phonons of the solid leads to a strong decrease of the relaxation time of the force acting on the liquid in axial direction.
 This is what leads to higher flux rates in flexible nanotubes.
Finally, relating the observations of this section to the main findings in the manuscript, the difference between predictions of the flow through CNTs made by classical force fields \cite{Falk2010} deviate even further from our ML-based simulations when we treat the walls as rigid.
This highlights the importance of describing the interatomic interactions at quantum-mechanical accuracy to obtain insight into the radius and material dependence in nanotubes.

\FloatBarrier
\newpage
%